*Submitted to PRB, revised version*

# Intrinsic structural instabilities of domain walls driven by gradient couplings: meandering anferrodistortive-ferroelectric domain walls in $BiFeO_3$

*Eugene A. Eliseev[1], Anna N. Morozovska[2,3], Christopher T. Nelson[4], and Sergei V. Kalinin[4*]*

*[1] Institute for Problems of Materials Science, National Academy of Sciences of Ukraine, Krjijanovskogo 3, 03142 Kyiv, Ukraine*

*[2] Institute of Physics, National Academy of Sciences of Ukraine, 46, pr. Nauky, 03028 Kyiv, Ukraine*

*[3] Bogolyubov Institute for Theoretical Physics, National Academy of Sciences of Ukraine, 14-b Metrolohichna str. 03680 Kyiv, Ukraine*

*[4] The Center for Nanophase Materials Sciences, Oak Ridge National Laboratory, Oak Ridge, TN 37831*

## Abstract

Using Landau-Ginzburg-Devonshire approach, we predict the intrinsic instability of the ferroelectric-ferroelastic domain walls in the multiferroic $BiFeO_3$ emerging from the interplay between the gradient terms of the antiferrodistortive and ferroelectric order parameters at the walls. These instabilities are the interface analogue of the structural instabilities in the vicinity of phase coexistence in the bulk; and so they do not steam from incomplete polarization screening in thin films or its spatial confinement, electrostrictive or flexoelectric coupling. The effect of $BiFeO_3$ material parameters on the 71°, 109°, and 180° walls is explored, and it is shown that the meandering instability appears at 109°, and 180° walls for small gradient energies, and the walls become straight and broaden for higher gradients. In contrast to the 180° and 109° domain walls, uncharged 71° walls are always straight, and their width increases with increasing the tilt gradient coefficient. The wall instability and associated intrinsic meandering provide a new insight into the behavior of morphotropic and relaxor materials, wall pinning, and mechanisms of interactions between order parameter fields and local microstructure.

[*] corresponding author, e-mail: sergei2@ornl.gov

## I. INTRODUCTION

Multiferroics, defined as materials with more than one ferroic long-range order [1, 2, 3], are ideal systems for fundamental studies of couplings among the order parameters of different nature, e.g. ferroelectric (**FE**) polarization, structural antiferrodistortion (**AFD**), ferromagnetic (**FM**) and antiferromagnetic (**AFM**) order parameters [4, 5, 6, 7, 8, 9, 10, 11]. The AFD, FE, FM, and AFM degrees of freedom in multiferroics are interlinked via different types of biquadratic couplings leading to versatile phase diagrams and complex domain structures [4-11]. In many cases, the interaction of the domain structures with underpinning frozen disorder gives rise to highly mobile structures and materials with giant functional responses.

The biquadratic couplings between AFD and other long-range orders are universal for all multiferroics with rotational antiferrodistortive symmetry [12]. The most common is the Houchmandazeh-Laizerowicz-Salje coupling, that is the biquadratic coupling between the AFD order parameter and FE polarization [13, 14]. AFD-FE coupling can significantly influence the structure and local properties of domain walls in AFD multiferroics [15, 16]. Similarly, the biquadratic magnetoelectric coupling, that is the coupling between polarization and magnetization [4, 5], can influence phase diagrams, domain wall structure and morphology [17]. The bilinear flexoelectric coupling [18], that coupled the strain gradient with polarization and vise versa, can induce incommensurate spatially modulated phases in ferroics including antiferroelectric (**AFE**) and AFD ones [19, 20, 21, 22]. The flexo-antiferrodistortive coupling, inherent to all AFD systems, can lead to the formation of incommensurate, spatially-modulated AFD and AFE phases in multiferroics [23], which are indeed observed in e.g. $Bi_ySm_{1-y}FeO_3$ [19], $EuTiO_3$ [24, 25]. There are also a wide variety of spatially modulated domain structures observed experimentally at the morphotropic boundaries in (multi)ferroics [26, 27, 28, 29, 30]

The vectorial nature of the AFD order parameter can strongly influence the phase stability, domain structure, polar, dielectric and magnetoelectric properties of (multi)ferroic thin films [31, 32, 33]. Sometimes phase diagrams of thin strained films are complicated by unusual low symmetry phases, which are absent in their bulk counterparts [34, 35, 36, 37]. Vortices and vertices composed by the closure of domain walls have been observed experimentally in nanoscale multiferroics [38, 39], especially in BFO [40, 41]. Fractal domain structures have been observed in multiferroic thin films [42] and near the surface of ferroelectric relaxors close to relaxor-ferroelectric transition [43].

Unusual polar structures with domain walls of labyrinthine shape (shortly "labyrinthine domain structure") have been observed near the surface of relaxors with so-called "ergodic phases" [44, 45] [46]. The labyrinthine domain structure was calculated theoretically in thin films of incommensurate and bi-layered ferroelectrics [47, 48], being similar to those observed in ultrathin magnetic films [49]. Spherical nanoparticles of uniaxial ferroelectrics $CuInP_2S_6$ and $Sn_2P_2S_6$ covered by a layer of screening charge with finite screening length revealed the transformation from a regular stripe domain structure into a labyrinthine one when the polarization gradient energy decreases below the critical value [50, 51]. The transformation can be identified as a gradient-driven morphological transition, and appeared unrelated with flexoelectric or electrostrictive, or any other bilinear, or biquadratic coupling influence.

To the best of our knowledge the physical origin of complex morphology of domain structures and modulated phases in nanoscale ferroics is the imbalance between domain wall surface energy and electrostatic or magnetic (or possibly elastic) energy contributions. Specifically, a ferroelectric nanoparticle tends to minimize its electrostatic energy by creation of the complex or/and irregular features of domain structure near the free surfaces, but the structure cannot be too fine-scale due to the increasing energy of domain walls (see e.g. discussion in Refs.[50-51]). Much more complex situation, corresponding to the balance of labyrinthine domain structure in the bulk and vortices at the surface, are expected in multiaxial ferroelectrics with polarization rotation allowed, such as $BaTiO_3$, $(Pb,Zr)TiO_3$ and $BiFeO_3$, and the fundamental question about the instability threshold of regular domain structure in nanoscale multiaxial multiferroics remains open.

The gap in the knowledge motivates this work that reveals a meandering zig-zag like instability of AFD-FE domain walls in thin BFO films. This unexpected result, obtained by finite element modeling (**FEM**), is explained within Landau-Ginzburg-Devonshire (**LGD**) theory framework.

The original part of the paper is structured as follows. LGD free energy is given in **section II.A**. The problem statement including the film geometry, brief form of the coupled Euler-Lagrange equations with boundary conditions are described in **section II.B**. The impact of biquadratic coupling on the stability of homogeneous phases is analyzed in **section II.C**. Simulation details with the special attention to the measures taken to establish the physical origin of the complex domain morphologies are described in **section III.A.** The appearance of low symmetry phases limited by $180^{\circ}$ or $109^{\circ}$ zig-zag like meandering AFD-FE domains and their changes with increasing of the gradient energy are presented in **section III.B** and **III.C,** respectively. The gradient-driven broadening of AFD-FE $71^{\circ}$ domain walls is discussed in **section III.D. Section IV** is a brief summary. Evident form of the free energy, boundary conditions and material parameters are given in **Appendix A** [52]. Supplementary figures are presented in **Appendix B** [52].

## II. THEORETICAL FORMALISM

As a model system, we have chosen bismuth ferrite $BiFeO_3$ (**BFO**) solid solutions. Pristine and rare-earth doped BFO is the unique multiferroic [53, 54] with a strong FE polarization, AFD oxygen octahedron rotation, FM and AFM long-range orders coexisting up to room and elevated temperatures. Specifically bulk BFO exhibits AFD long-range order at temperatures below 1200 K; it is FE with a large spontaneous polarization below 1100 K and AFM below Neel temperature $T_N \approx$ 650 K [55]. Notably that the behavior of the AFD order parameter at the BFO domain walls determines their structure and energy [56]. Domain walls in BFO exhibit unusual electrophysical properties, such as conduction and magnetotransport enhancement [57, 58, 59, 60, 61, 62]. Recently, a complete phase diagram of BFO including the AFM, FE, and AFD phases was calculated within LGD theory [63].

The pronounced multiferroic properties and unusual domain structure evolution maintain in BFO thin films and heterostructures [64, 65, 66, 67, 68, 69, 70, 71, 72]. In particular, atomic mapping of structural distortions in 109° domains revealed that the coexistence of rhombohedral and orthorhombic phases in ultrathin BFO films can be driven by interfacial oxygen octahedral coupling [73, 74]. The role of the rotomagnetic coupling, that is the biquadratic coupling between the AFD and AFM (or FM) orders [75], has been studied in BFO fine grained ceramics [76].

### A. Landau-Ginzburg-Devonshire free energy

Thermodynamic LGD potential $G$ that describes AFD, FE and AFM properties of BFO is:

$$G=\int_V\left(\Delta G_{AFD}+\Delta G_{FE}+\Delta G_{AFM}+\Delta G_{BQC}+\Delta G_{\sigma}\right)dv+\int_S\left(\Delta G_{AFD}+\Delta G_{FE}\right)dS . \quad (1)$$

The AFD energy $\Delta G_{AFD}$, corresponding to $R3c$ phase, is a six-order expansion on the oxygen tilt $\Phi_i$ and its gradients,

$$\Delta G_{AFD}=b_i\,(T)\Phi_i^2+b_{ij}\Phi_i^2\Phi_j^2+b_{ijk}\Phi_i^2\Phi_j^2\Phi_k^2+v_{ijkl}\frac{\partial\Phi_i}{\partial x_k}\frac{\partial\Phi_j}{\partial x_l} . \quad (2)$$

Here $\Phi_i$ are components of the pseudo-vector determining the out-of-phase static rotations of the oxygen octahedral groups , and Einstein summation convention is employed. In accordance with the classical LGD theory, we assume that the coefficients $b_i$ are temperature dependent in accordance with the Barrett law [77], $b_i=b_T T_{q\Phi}\left(\coth\left(T_{q\Phi}/T\right)-\coth\left(T_{q\Phi}/T_{\Phi}\right)\right)$, where $T_{\Phi}$ is the AFD transition temperature, $T_{q\Phi}$ is a Barrett temperature [78]. Other coefficients in Eqs.(2) are temperature independent.

FE energy $\Delta G_{FE}$ is a six-order expansion on the polarization vector $P_i$ and its gradients,

$$\Delta G_{FE}=a_i\,(T)P_i^2+a_{ij}P_i^2P_j^2+a_{ijk}P_i^2P_j^2P_k^2+g_{ijkl}\frac{\partial P_i}{\partial x_k}\frac{\partial P_j}{\partial x_l}-P_iE_i-\frac{\varepsilon_0\varepsilon_b}{2}E_i^2 . \quad (3)$$

The coefficients $a_k$ are temperature dependent, $a_k^{(P)} = \alpha_T \left(T_{qP} \coth\left(T_{qP}/T\right) - T_C\right)$, where $T_C$ is a Curie temperature, $T_{qP}$ is the characteristic Barrett temperature related with some "vibrational modes" [77]. Other coefficients in Eqs.(3) are temperature independent. $E_i$ are the components of internal electric field related with electrostatic potential $\varphi$ in a standard way $E_i = -\partial\varphi/\partial x_i$. Universal dielectric constant is $\varepsilon_0$, $\varepsilon_b$ is the dielectric permittivity of background [18, 79].

AFM energy $\Delta G_{AFM}$ is a fourth-order expansion in terms of the AFM order parameter vector $L_i$ and its gradient, as follows from the fact that this phase transition in BFO is second order [63]. The details of $\Delta G_{AFM}$ is considered elsewhere [63].

The AFD-FE coupling energy $\Delta G_{BQC}$ is the biquadratic function of $P_i$ and $\Phi_i$:

$$\Delta G_{BQC} = \zeta_{ijkl} \Phi_i \Phi_j P_k P_l \ . \tag{4}$$

The temperature-independent coefficients $\zeta_{ijkl}$ are the components of AFD-FE biquadratic coupling tensor.

The elastic, electrostriction, rotostriction and flexoelectric energy is

$$\Delta G_{\sigma} = -s_{ijkl}\sigma_{ij}\sigma_{kl} - Q_{ijkl}\sigma_{ij} P_k P_l - R_{ijkl}\sigma_{ij}\Phi_k\Phi_l - \frac{F_{ijkl}}{2}\left(\sigma_{ij}\frac{\partial P_k}{\partial x_l} - P_k \frac{\partial \sigma_{ij}}{\partial x_l}\right). \tag{5}$$

Here $s_{ijkl}$ are the components of elastic compliances tensor, $Q_{ijkl}$ are the components of electrostriction tensor, $R_{ijkl}$ are the components of rotostriction tensor, and $F_{ijkl}$ are the components of flexoelectric tensor.

The surface energy has the form:

$$\int_S \left(\Delta G_{AFD} + \Delta G_{FE}\right) dS = \int_S \left(\frac{b_i^{(S)}}{2}\Phi_i^2 + \frac{a_i^{(S)}}{2} P_i^2\right) dS \tag{6}$$

Surface energy coefficients $b_i^{(S)}$ and $a_i^{(S)}$ have different nature and control the broadening of ADF and FE domain walls at the surface, respectively.

### B. Problem statement

Let us consider a BFO film of thickness *h* placed in a perfect electric contact with conducting bottom electrode that mechanically clamps the film. The top surface of the film is mechanically free and can be in an ideal electric contact with the top electrode, or electrically open, or covered with the surface screening charge. The charge density $\sigma(\varphi)$, appearing due to surface states [80], or electro-chemically active ions [81, 82, 83, 84], depends on the electric potential φ [see **Fig. 1(a)**]. **Figures 1(b)-(d)** show three types of nominally uncharged 180°, 109° and 71° domain walls in BFO.

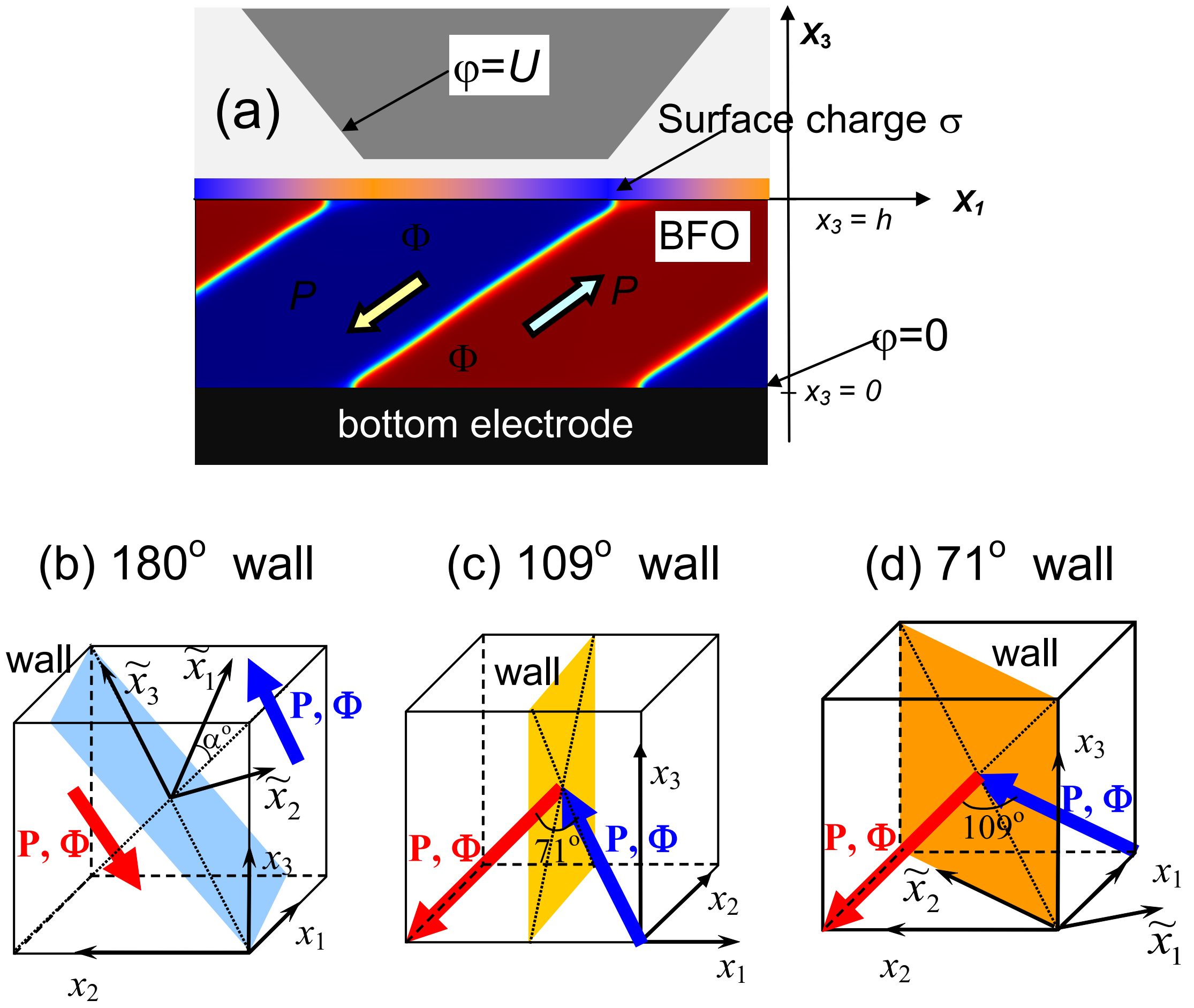

**FIGURE. 1**. **(a)** Considered system, consisting of electrically conducting bottom electrode, BFO film of thickness $h$ with a domain structure, surface screening charge with density $\sigma(\varphi)$ and ambient media (from bottom to the top). Three types of nominally uncharged 180 ° **(b)**, 109 ° **(c)** and 71 ° **(d)** domain walls in BFO are shown in the bottom row.

Electrostatic potential inside the ferroelectric film satisfies the Poisson equation, $\varepsilon_0\varepsilon_b\Delta\varphi - \operatorname{div}\vec{P} = 0$, and Laplace equation is valid in the dielectric gap, i.e. $\varepsilon_0\varepsilon_e\Delta\varphi = 0$ ($\varepsilon_e$ is the dielectric permittivity of external media). Electric boundary conditions are zero electric potential at the bottom of the film contacting the conducting substrate, $\left.\varphi\right|_{x_3=0} = 0$, and the potential continuity, $\left.\varphi\right|_{x_3=h-0} - \left.\varphi\right|_{x_3=h+0} = 0$, at the interface between the ferroelectric film and the ambient medium. Another boundary condition at interface $x_3 = h$ is for the normal components of the electric displacement, namely $D_3\big|_{x_3=h+0} - D_3\big|_{x_3=h-0} = \sigma(\varphi)\big|_{x_3=h}$ where $D_3 = P_3 - \varepsilon_0\varepsilon_b \dfrac{\partial\varphi}{\partial x_3}$ in the film, ($0 < x_3 < h$) and $D_3 = -\varepsilon_0\varepsilon_e \dfrac{\partial\varphi}{\partial x_3}$ in the dielectric gap ($h < x_3$). Here, we consider the special case of the surface screening charge with the density given by expression, $\sigma(\varphi) = -\varepsilon_0\varphi/\Lambda$, where $\Lambda$ is the effective screening length [85, 86]. Typically the value of $\Lambda$ is smaller or even significantly smaller than 1 nm [87, 88]. The condition $\Lambda \to 0$ corresponds to the perfect electric contact between the top conducting electrode and the film [89], and we consider the limit for comparison. The top electrode can be either biased ($\varphi\big|_{x_3=h} = U$) or grounded ($\varphi\big|_{x_3=h} = 0$), depending on the experimental situation corresponding to the SPM tip placed on the film surface.

Elastic problem formulation is based on the modified Hooke's law obtained using the thermodynamic relation $u_{ij} = -\dfrac{\delta G_{ELS}}{\delta\sigma_{kl}}$, where $u_{ij}$ are the components of elastic strain tensor. Mechanical equilibrium conditions are $\partial\sigma_{ij}/\partial x_j = 0$ [90]. Note that the film-substrate interface was considered as unstrained one (misfit strain is zero) corresponding to the elastically matched substrate.

The system of coupled Euler-Lagrange equations allowing for Khalatnikov relaxation of the oxygen tilt and polarization components, $\Phi_i$ and $P_i$ , is:

$$\frac{\delta G}{\delta P_i} = -\Gamma_P \frac{\partial P_i}{\partial t} \quad \text{and} \quad \frac{\delta G}{\delta \Phi_i} = -\Gamma_\Phi \frac{\partial \Phi_i}{\partial t}. \tag{7a}$$

These equations are supplemented by the boundary conditions of zero generalized fluxes at the film boundaries,

$$\left. b^{(S)}\Phi_i + v_{ijkl}\frac{\partial \Phi_j}{\partial x_k} n_l \right|_S = 0, \quad \left. a^{(S)}P_i + g_{ijkl}\frac{\partial P_j}{\partial x_k} n_l \right|_S = 0 \quad (i=1, 2, 3). \tag{7b}$$

The so-called "natural boundary conditions" for the tilt, $\left. v_{ijkl}\dfrac{\partial \Phi_j}{\partial x_k} n_l \right|_S = 0$, correspond to $b^{(S)} = 0$; and the natural boundary conditions for polarization, $\left. g_{ijkl}\dfrac{\partial P_j}{\partial x_k} n_l \right|_S = 0$, corresponds to $a^{(S)} = 0$ in Eq.(7b). The natural conditions, which will be used hereinafter, correspond to the absence of surface energy (6) [91]. The explicit form of the free energy (1)-(6), Euler-Lagrange equations (7a) with boundary conditions (7b) are listed in **Appendix A** [52].

### C. The impact of biquadratic coupling on the stability of homogeneous R3c phase

Experimentally, bulk BFO should be in a rhombohedral R3c phase at temperatures below $T_C$. Since the biquadratic coupling and gradient energy coefficients in the free energy (1)-(5) are poorly known, one should be very careful with the choice of their numerical values in order to prevent the appearance of so-called nonphysical "extra" phases [63], which do not exist in reality and should be eliminated from the theoretical analysis of the domain structure configuration. Therefore, priory to study the effect of the gradient energy on domain structure, let us analyze whether any extra phase can be (meta)stable below $T_C$ for a chosen free energy functional form (1)-(5) with parameters taken from Refs. [63, 76] and listed in **Table I**. For this purpose let us perform the following analytical and numerical calculations.

Without biquadratic coupling contribution, i.e. for $\Delta G_{BQC} = 0$, and neglecting the 6-th powers of the polarizations and tilts, and their gradients, the energies of oxygen tilts and polarization are decoupled, and, using the idea of Dzyaloshinsky substitution [92], one can introduce the new variables $\Phi^2 = \frac{\Phi_1^2 + \Phi_2^2 + \Phi_3^2}{\sqrt{3}}$, $\Psi^2 = \frac{\Phi_3^2 - \Phi_1^2}{\sqrt{2}}$ and $\Omega^2 = \frac{2\Phi_2^2 - \Phi_1^2 - \Phi_3^2}{\sqrt{6}}$, which diagonalize the AFD contribution to the free energy. Similar substitution for polarization components, $P^2 = \frac{P_1^2 + P_2^2 + P_3^2}{\sqrt{3}}$, $Q^2 = \frac{P_3^2 - P_1^2}{\sqrt{2}}$ and $R^2 = \frac{2P_2^2 - P_1^2 - P_3^2}{\sqrt{6}}$, diagonalizes the FE energy. Namely:

$$\Delta G_{AFD}^{2-4}[\Phi, \Psi, \Omega] = \sqrt{3} b_1 \Phi^2 + (b_{11} + b_{12})\Phi^4 + \left(b_{11} - \frac{b_{12}}{2}\right)\left(\Psi^4 + \Omega^4\right), \tag{8a}$$

$$\Delta G_{FE}^{2-4}[P, Q, R] = \sqrt{3} a_1 P^2 + (a_{11} + a_{12})P^4 + \left(a_{11} - \frac{a_{12}}{2}\right)\left(Q^4 + R^4\right). \tag{8b}$$

Expressions (8) have the four global equivalent minimums in the AFD-FE phase, which are stable at $b_1 < 0$, $b_{11} - \frac{b_{12}}{2} > 0$, $a_1 < 0$, $a_{11} - \frac{a_{12}}{2} > 0$. The coordinates of the minimums in the six dimensional (6D) phase space are

$$\{\Phi, \Psi, \Omega, P, Q, R\} = \left\{ \pm\sqrt{-\frac{\sqrt{3} b_1}{2(b_{11} + b_{12})}}, 0, 0, \pm\sqrt{-\frac{\sqrt{3} a_1}{2(a_{11} + a_{12})}}, 0, 0 \right\} \tag{9}$$

.Each of the minimums correspond to the conventional R3c phase of BFO, in which $\Phi_1^2 = \Phi_2^2 = \Phi_3^2 = -\frac{b_1}{2(b_{11} + b_{12})}$ and $P_1^2 = P_2^2 = P_3^2 = -\frac{a_1}{2(a_{11} + a_{12})}$.

Nonzero biquadratic coupling energy $\zeta_{ijkl}\Phi_i\Phi_j P_k P_l$ given by Eq.(4), as well as 6-th order powers $\Phi_i^6$ and $P_i^6$ included in Eqs.(2)-(3), make the diagonalization (8) impossible. The minimums can be shifted, and, moreover, some of them can become metastable or even disappear due to the biquadratic coupling and 6-th order terms contribution. Specifically, in coordinates $\{\Phi, \Psi, \Omega, P, Q, R\}$ the "isotropic" part of biquadratic energy $\Delta G_{BQC}^{11}$ can be identically rewritten as:

$$\Delta G_{BQC}^{11} = \zeta_{11}\left(\Phi_1^2 P_1^2 + \Phi_2^2 P_2^2 + \Phi_3^2 P_3^2\right) \equiv \zeta_{11}\left(\Phi^2 P^2 + \Psi^2 Q^2 + \Omega^2 R^2\right) \qquad (10)$$

The oversimplified free energy (7)-(8) including the isotropic biquadratic coupling energy (10) has the form

$$\Delta G_{AFD-FE}^{2-4} = \begin{bmatrix} \sqrt{3}b_1\Phi^2 + (b_{11}+b_{12})\Phi^4 + \left(b_{11} - \frac{b_{12}}{2}\right)\left(\Psi^4 + \Omega^4\right) + \sqrt{3}a_1 P^2 + \\ (a_{11}+a_{12})P^4 + \left(a_{11} - \frac{a_{12}}{2}\right)\left(Q^4 + R^4\right) + \zeta_{11}\left(\Phi^2 P^2 + \Psi^2 Q^2 + \Omega^2 R^2\right) \end{bmatrix} \qquad (11)$$

The energy (11) has four energetically equivalent minimums with coordinates

$$\{\Phi, \Psi, \Omega, P, Q, R\} = \left\{ \pm\sqrt{\frac{\sqrt{3}\left[a_1\zeta_{11} - 2b_1(a_{11}+a_{12})\right]}{4(b_{11}+b_{12})(a_{11}+a_{12}) - \zeta_{11}^2}},\ 0,\ 0, \pm\sqrt{\frac{\sqrt{3}\left[b_1\zeta_{11} - 2a_1(b_{11}+b_{12})\right]}{4(a_{11}+a_{12})(b_{11}+b_{12}) - \zeta_{11}^2}},\ 0,\ 0 \right\} \qquad (12)$$

These minimums correspond to R3c phase in a bulk AFD-FE multiferroic with isotropic biquadratic coupling. Unfortunately, we could not find any analytical expressions for the minimum coordinates if the anisotropic biquadratic coupling (4) is included in the free energy (1).

Calculations were performed for the 2-4-6 coupled free AFD-FE energy (1)-(5) with BFO parameters from **Table I**. Cross-sections of the free energy surface have been calculated without [$\zeta_{ijkl} = 0$, **Fig.2(a)**] and with [$\zeta_{ijkl} \neq 0$, **Fig.2(b)**] biquadratic coupling energy (4)**.** The differences between **Fig.2(a)** and **2(b)** are caused by sotropic terms $\zeta_{11}\left(\Phi_1^2 P_1^2 + \Phi_2^2 P_2^2 + \Phi_3^2 P_3^2\right)$ in Eq.(10) and anisotropic terms $\zeta_{44}\left(\Phi_1\Phi_2 P_1 P_2 + \Phi_2\Phi_3 P_2 P_3 + \Phi_1\Phi_3 P_1 P_3\right)$ in Eq.(4). Four equivalent deepest minimums with nonzero coordinates $\Phi_1 = \Phi_2 = \Phi_3 \neq 0$ and $P_1 = P_2 = P_3 \neq 0$, and zero coordinates $\Psi = \Omega = Q = R = 0$ are seen in **Fig.2(a)-2(b)** [see also Eq.(9)]. The minimums are separated by a local maximum at the coordinate origin and four saddle points. The case $\Phi_1^2 = \Phi_2^2 = \Phi_3^2$, $P_1^2 = P_2^2 = P_3^2$ and $\Psi = \Omega = Q = R = 0$ corresponds to the stable R3c phase.

The free energy dependence on polarization at fixed tilt components $\Phi_i = \Phi_1 = \Phi_2 = \Phi_3 = 22$ pm, and the dependence of the energy on the tilt at fixed polarization components $P_i = P_1 = P_2 = P_3 = 0.48$ C/m² are shown in **Figs.2(c)** and **2(d),** respectively. The influence of the coupling makes the minimums deeper, but does not shift or eliminate them [compare solid and dashed curves in **Fig.2 (c)** and **2(d)**].

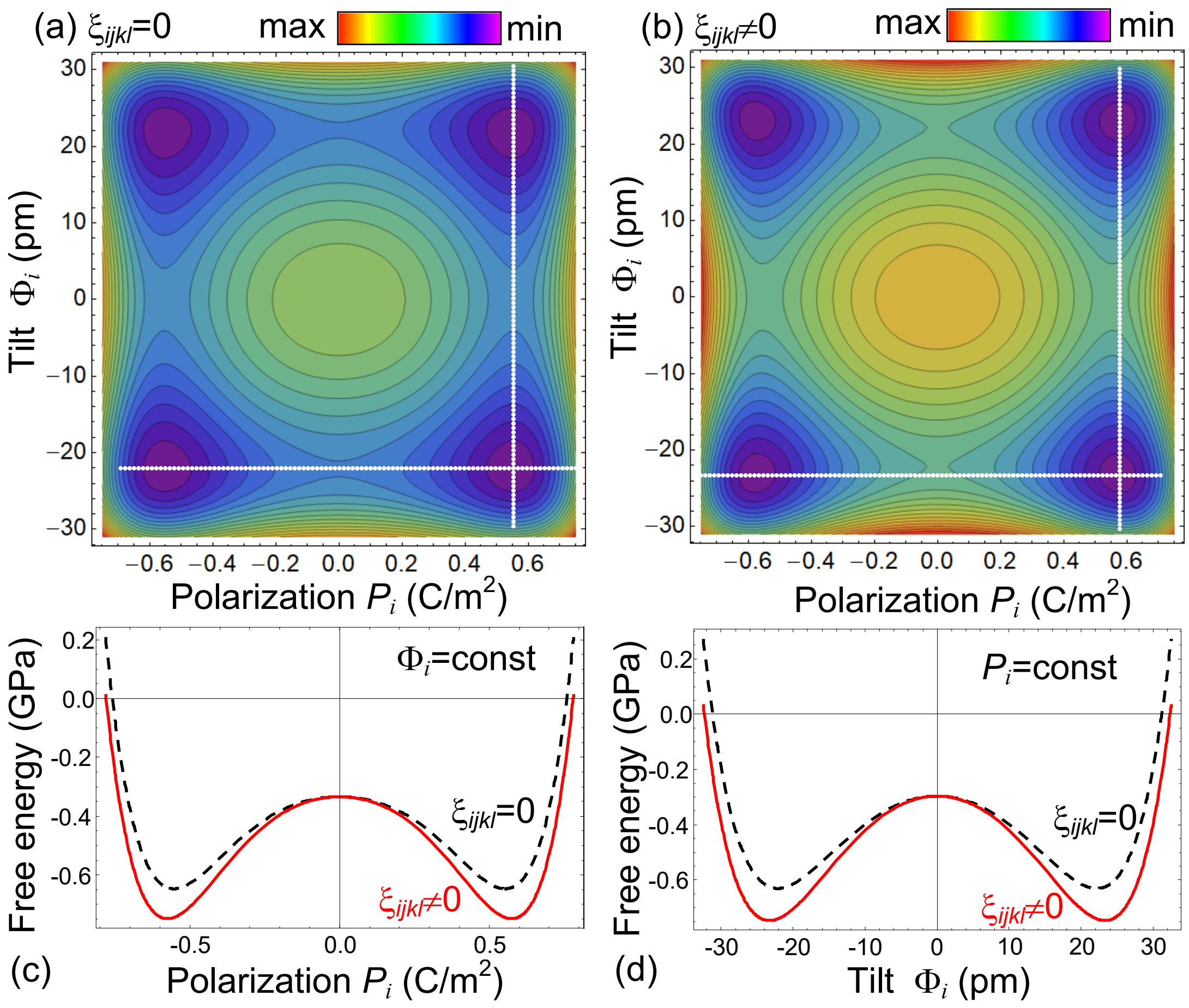

**FIGURE. 2**. Free energy dependence on the tilt and polarization components, $\Phi_i = \Phi_1 = \Phi_2 = \Phi_3$ and $P_i = P_1 = P_2 = P_3$. Contour maps were calculated at room temperature without **(a)** and with **(b)** biquadratic coupling contribution. The free energy dependence on $\Phi_i$ at fixed $P_i = 0.48$ C/m² **(c)**; and its dependence on $P_i$ at fixed $\Phi_i = 22$ pm **(d)**. Dashed and solid curves in plots **(c, d)** show the cases without ($\zeta_{ijkl} = 0$) and with ($\zeta_{ijkl} \neq 0$) biquadratic coupling contribution, respectively. Corresponding cross-sections are shown by dotted lines in plots **(a)** and **(b)**. BFO parameters are listed in **Table I,** T=300 K**.**

We checked numerically, that any other local (or global) minima corresponding to nonzero coordinates $\{P, \Phi, \Psi, \Omega, Q, R\}$ is absent for the BFO parameters listed in **Table I**. Hence among all homogeneous phases, only R3c phase is absolutely stable below $T_C$ in a bulk BFO without polarization or tilt gradient energy. The result is expected and confirms the appropriate choice of the free energy form given by Eqs.(1)-(5) and numerical parameters in **Table I**. That is why we can conclude that any other metastable or stable phases or/and domain configurations, different from R3c, which will be revealed and analyzed in the next sections, cannot originate from the "extra" local minima of the free energy (1)-(5).

**Table I.** BFO parameters used in LGD calculations (taken from Refs. [63, 76])

| **Parameter** | **Designation** | **Numerical value for BFO** |
|---|---|---|
| background permittivity | $\varepsilon_b$ | 7 |
| dielectric stiffness | $\alpha_T$ ($\times10^5$C$^{-2}\cdot$Jm/K) | 9 |
| Curie temperature for P | $T_C$ (K) | 1300 |
| Barrett temperature for P | $T_{qP}$ (K) | 800 |
| 4th order coefficients in the polarization expansion | $a_{ij}$ ($\times10^8$C$^{-4}\cdot$m$^5$J) | $a_{11}$= –13.5, $a_{12}$= 5 |
| 6th order coefficients in the polarization expansion | $a_{ijk}$ ($\times10^9$C$^{-6}\cdot$m$^9$J) | $a_{111}$= 11.2, $a_{112}$= -3, $a_{123}$= –6 |
| electrostriction | $Q_{ij}$ (C$^{-2}\cdot$m$^4$) | $Q_{11}$=0.03, $Q_{12}$= –0.01, $Q_{44}$=0.01 |
| rotostriction | $R_{ij}$ ($\times10^{18}$ m$^{-2}$) | $R_{11}$= –1.32, $R_{12}$= –0.43, $R_{44}$=8.45 |
| compliances | $s_{ij}$ ($\times10^{-12}$ Pa$^{-1}$) | $s_{11}$=8.3, $s_{12}$= –2.7, $s_{44}$=9.25 |
| polarization gradient coefficients | $g_{ij}$ ($\times10^{-10}$C$^{-2}$m$^3$J) | $g_{11}$=5, $g_{12}$= –0.5, $g_{44}$=0.5 |
| AFD-FE coupling | $\times10^{29}$ C$^{-2}\cdot$m$^{-2}$ J/K | $\xi_{11}$ = –0.5, $\xi_{12}$ =0.5, $\xi_{44}$ = –2.6 |
| 2nd order coefficients in the tilt expansion | $b_T$($\times10^{26}\cdot$J/(m$^5$K)) | 4 |
| Curie temperature for Φ | $T_\Phi$ (K) | 1440 |
| Barrett temperature for Φ | $T_{q\Phi}$ (K) | 400 |
| 4th order coefficients in the tilt expansion | $b_{ij}$ ($\times10^{48}$J/(m$^7$)) | $b_{11}$= –24+4.5 $\left(\coth(300/T)-\coth(3/14)\right)$ <br> $b_{12}$= 45–4.5 $\left(\coth(300/T)-\coth(1/4)\right)$ |
| 6th order coefficients in the tilt expansion | $b_{ijk}$ ($\times10^{70}$ J/(m$^9$)) | $b_{111}$= 4.5–3.4 $\left(\coth(400/T)-\coth(2/7)\right)$ <br> $b_{112}$= 3.6–0.04 $\left(\coth(10/T)-\coth(1/130)\right)$ <br> $b_{123}$= 41–43.2 $\left(\coth(1200/T)-\coth(12/11)\right)$ |
| tilt gradient coefficients | $v_{ij}$ ($\times10^{11}$ J/ m$^3$) | $v_{11}$=0.25, $v_{44}$=(0.25 – 25) |
| polarization extrapolation length | $\lambda_i^P \equiv g_{i3}^{(P)}/a_i^{(S)}$ (nm) | Varied from zero to high h values > 100 nm |
| tilt extrapolation length | $\lambda_i^\Phi \equiv v_{i3}^{(\Phi)}/b_i^{(S)}$ (nm) | Varied from zero to high h values > 100 nm |
| effective screening length | Λ (nm) | Varied from zero to 0.1 nm. Λ=0 in Figs.3 – 12. |

The values of the flexoelectric tensor components $F_{ijkl}$ are not listed in **Table I** due to their small impact on the studies phenomena. We vary them within typical range $0 \leq |F_{ijkl}| \leq 10^{11}$ m³/C.

## III. SIMULATION RESULTS AND DISCUSSION

### A. Simulation details

We used FEM to simulate the oxygen tilt and polarization distributions in thin free-standing BFO films covered by conducting electrodes. The film thickness $h$ varied from 5 nm to 500 nm, and the typical picture of domain morphology was observed at $h$>15 nm; so we use the thicknesses (16 – 20) nm for illustration. BFO parameters used in the FEM calculations to generate figures are listed in **Table I.**

The values of $a^{(S)}$ and $b^{(S)}$ in the boundary conditions (7b) significantly affect on the "**critical thickness**" of the film [91]. The critical thickness is the thickness below which the size-induced phase transition to a parent (e.g. paraelectric) phase without long-range order occurs (see e.g. Refs.[91, 93]). As it follows from the analytical expressions derived in Refs.[47, 91] the critical thickness of FE (or AFD) transition decreases with $a^{(S)}$ (or $b^{(S)}$) increase. So that the application of the natural boundary conditions ($a^{(S)} = b^{(S)} = 0$) leads to the minimal critical thickness of the film (see e.g. [47, 91]).

To minimize the influence of the surface on obtained results, we put $a^{(S)} = b^{(S)} = 0$ in the boundary conditions (7b). For comparison we performed simulations for zero polarization and tilt components at the film surfaces, $P_i \big|_{x_3=0,h} = 0$ and $\Phi_i \big|_{x_3=0,h} = 0$, which corresponds to the maximal influence of the surface. These conditions lead to the maximal critical thickness of the film [47, 91].

It appeared that curved walls arise as a result of the relaxation process of a random domain distribution, named "random seeding" (see **Fig. 3**). Also the random seeding can be superimposed on the ideal nominally uncharged 180°, 109° and 71° domain wall structure in R3c phase. Initial and final domain states are shown in **Fig.4** and **Fig.S2** in **Appendix B** [52].

From **Figs.3, 4** and **S2** the "curved", "meandering" and "zig-zag" like features appeared at the 180° and 109° AFD walls, but not at the 71° walls. Schematic images of the straight, curved, meandering and zig-zag like domain wall profiles are shown in **Fig.S1** in **Appendix B** [52].

The energy density excess $\Delta G$ corresponding to the relaxation of the initial random domain distribution (**RD**), poly-domain distribution (**PD**) with straight 180°, 109° and 71° domain walls; and poly-domain distribution disturbed by a random seeding (**PD + RD**) have been compared. It appeared that the energies are surprisingly close, namely $\Delta G = -19.926$ J/m$^2$ for the curved domain walls obtained from the relaxation of RD (see **Fig. 3**), $\Delta G = -19.853$ J/m$^2$ for the 180° domain walls obtained from the relaxation of PD+RD [see **Fig. 4(a)-(d)**], and $\Delta G = -19.865$ J/m$^2$ for the 180° domain walls obtained from the relaxation of PD [see **Fig. 4(e)-(f)**]. The final distributions of the polarization and tilt shown in **Figs. 3-4,** which have high negative and approximately equal energies $\Delta G \approx -19.9$ J/m$^2$, are long-living metastable states of the curved domain walls in BFO [94].

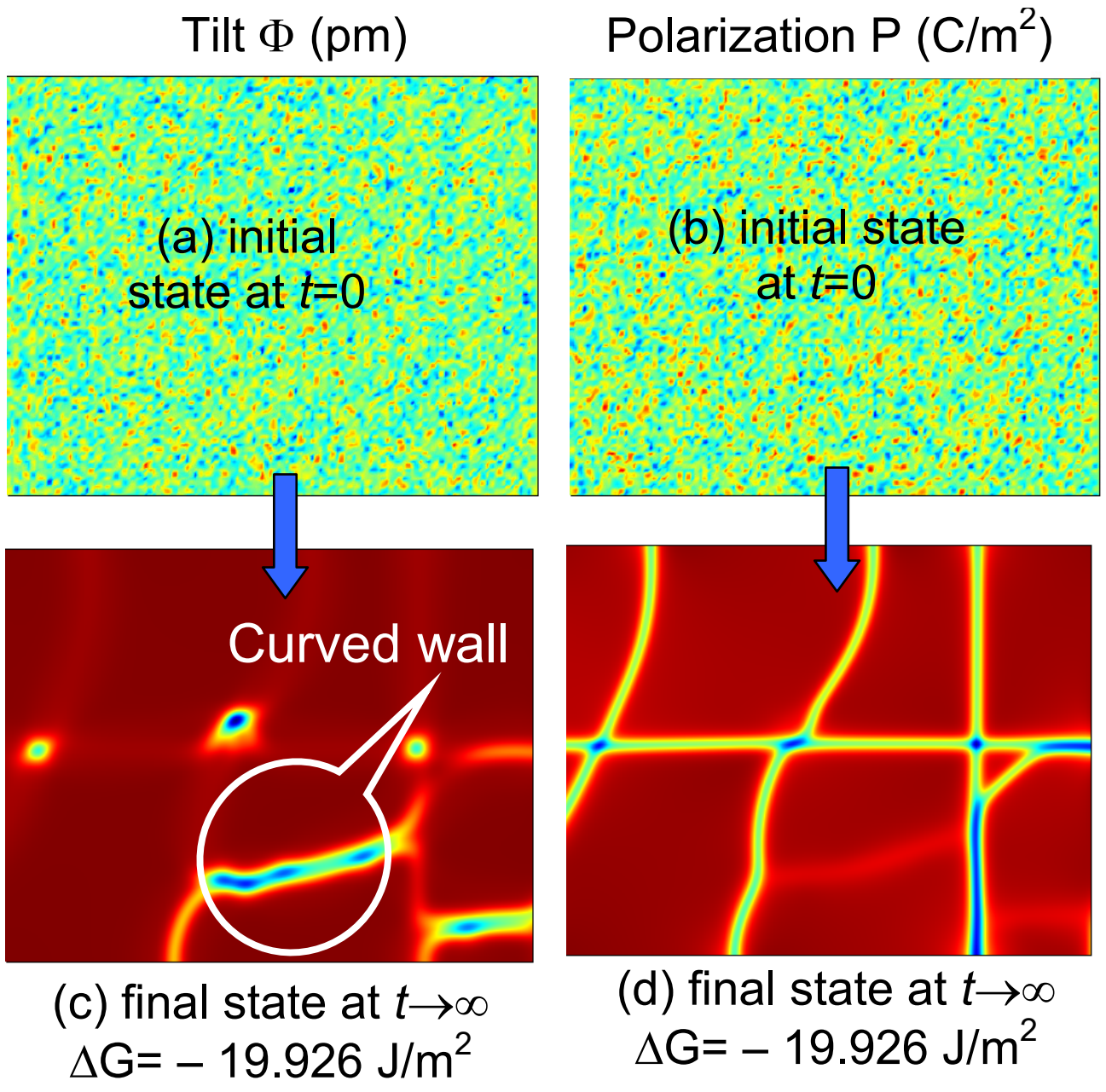

**FIGURE 3.** Initial random seeding and final distributions of the tilt **(a, c)** and polarization **(b,d)** calculated in a 16-nm BFO film. Gradient coefficient $v_{44}$=0.25×10$^{11}$ J/ m$^3$, T=300 K, other parameters are listed in **Table I.**

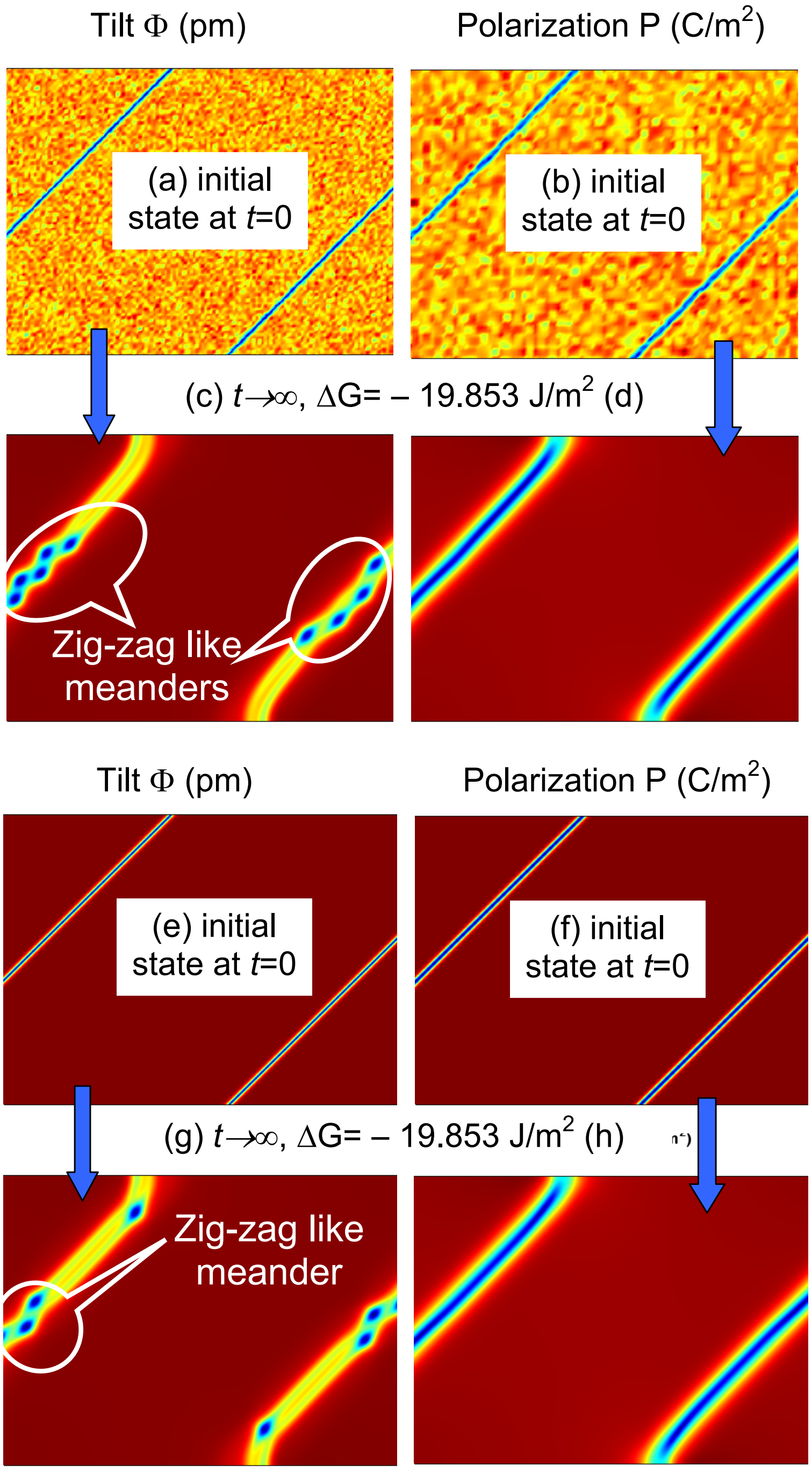

**FIGURE 4.** Initial and final distributions of the tilt **(a, c)** and polarization **(b,d)** calculated in a 16-nm BFO film. The 180° domains were imposed on random seeding. Gradient coefficient $v_{44}$=0.25×10$^{11}$ J/m$^3$, T=300 K, other parameters are listed in **Table I.**

From the analysis of the free energy relief presented in **section II.C** we cannot establish the origin of the curved, meandering and zig-zag like AFD-FE walls shown in **Fig.3(c,d)** and **Fig.4(c,g)**. So, what is the physical origin of the meandering and zig-zag like 180° and 109° domain boundaries? Why the effect does not exist for at 71° domain walls? To establish the origin, we performed the following numerical experiments.

(a) The unusual meandering AFD-FE domain structures exist and become insensitive to the screening length values at $\Lambda << 0.1$ nm. We observed the stable meandering domains in the limiting case $\Lambda \to 0$ corresponding to perfect screening and minimal depolarization electric field. Hence the origin of the meandering walls is not the incomplete screening of ferroelectric polarization by the imperfect electrodes or surface charge. At that the low symmetry domains limited by the meandering walls is quite possible.

(b) We explored whether the meandering walls originate from the spatial confinement of polarization components at the film surfaces. Namely we compared the changes of the AFD-FE domain morphology when the polarization extrapolation lengths $\lambda_i^P \equiv \frac{g_{i3}^{(P)}}{a_i^{(S)}}$ vary from 0 (corresponding to zero polarization at the film surfaces, $P_i\big|_{x_3=0,h} = 0$) to infinity (corresponding to $a_i^{(S)} = 0$ and $\left.\frac{\partial P_i}{\partial x_3}\right|_{x_3=0,h} = 0$). For $\lambda_i^P = 0$ we see the appearance of FE domain wall broadening at the surface and its gradual decrease with $\lambda_i^P$ increasing , as anticipated. However no significant changes of the meandering walls occur in the film with $\lambda_i^P$ changing.

(c) We further compared the changes of the AFD-FE domain structure when the tilt extrapolation lengths $\lambda_i^\Phi \equiv \frac{v_{i3}^{(\Phi)}}{b_i^{(S)}}$ vary from 0 (corresponding to zero polarization at the film surfaces, $\Phi_i\big|_{x_3=0,h} = 0$) to the infinity (corresponding to $b_i^{(S)} = 0$ and $\left.\frac{\partial \Phi_i}{\partial x_3}\right|_{x_3=0,h} = 0$). For $\lambda_i^\Phi = 0$ we see the appearance of AFD domain wall broadening at the surface and its gradual decrease with $\lambda_i^\Phi$ increasing. However, no significant changes of the domain morphology occur with $\lambda_i^\Phi$ changing. Further we can assume that the spatial confinement delineates the appropriate boundary conditions for the oxygen tilt and polarization components at the film surfaces.

(d) We rotate the film surface cut to find the angle for which both AFD and FE walls are straight without inclusion of any other phases. We made sure that the angle does not exist. Also, we checked whether the meandering walls originate from the spatial confinement effect delineated by the appropriate boundary conditions for the oxygen tilt at the film surfaces. We increase the film thickness up to 500 nm and see that no significant changes in the morphology of meandering domains occur.

(e) Finally, we varied the components of the electrostrictive and flexoelectric couplings tensors in a typical range ($0 \le |F_{ijkl}| \le 10^{11}$ m$^3$/C, $0 \le |Q_{ijkl}| \le 0.1$ m$^4$/C$^2$), and lead to the conclusion that the appearance of observed effects do not steam from the couplings, because the meandering AFD walls weakly react on the changes of $F_{ijkl}$ and $Q_{ijkl}$ values.

(f) It appeared that the change of the tilt gradient coefficients $v_{ijkl}$ significantly affects on the curvature and meandering of domain walls, including the monoclinic phase appearance at the curved walls. It is important to underline that the monoclinic phase can be stable in a ferroic with one vectorial long-range order parameter, e.g. in a "normal" ferroelectric with a polarization vector **P**, if the higher-order powers of **P** (from 8$^{th}$ to 12$^{th}$) are included into the LGD free energy [95]. As a matter of fact, we consider two vectors, **P** and **Φ**, as the long-range order parameters, using 2-4-6 LGD expansion for each of them in numerical modeling [see Eqs.(2)-(3) and **Table I**], and simplified 2-4 expansion in analytical calculations [see Eqs.(8)]. Hence the effective order of the nonlinearity in the coupled Euler-Lagrange equations for **P** and **Φ** is 12 for numerical calculations and 8 for analytical calculations, making the appearance of monoclinic phases quite possible. Earlier we have found that the monoclinic phase can be stabilized without 8$^{th}$ powers of polarization, if the coupling between **P** and **Φ** is included [34].

(e) The impact of the polarization gradient coefficients $g_{ijkl}$ is much less pronounced, because the FE walls do not bend in order to remain uncharged. The charging of FE wall by the polarization bound charge will immediately lead to the appearance of strong depolarization electric field $E_i^d$ ($div\vec{E}^d \sim -div\vec{P}$) that's energy excess $-E_i^d P_i/2$ is positive at the region of the curved wall and relatively high. Thus, the polarization sub-system behaves in such a way to prevent the charging.

From the analysis of (a)-(e) we concluded that the origin of meandering AFD-FE domain walls is the coupling between the tilts and polarization gradient coefficients. This conclusion is consistent with the results of Conti et al. [96] for multiferroics with symmetric free energy and two order parameters. Conti et al. used a simple phenomenological model and have shown that the maximum and minimum near the antiphase domain walls appear on the profile of one of the order parameters depending on the anisotropy gradient energy, in the mixed phase when both order parameters are nonzero. Despite the fact that we consider a much more complex system with six order parameters, the extremums observed near the domain walls are qualitatively similar to the ones predicted by Conti et al. Thus, the appearance of maxima and minima on the profiles of the order parameters near the domain walls can be associated with the features (such as anisotropy) of the gradient energy.

To quantify the statement, one can introduce the tilt correlation length $L_C^{\Phi} \sim \sqrt{v}$ that is defined from the correlation function of the tilt vector fluctuations. The correlation length of polarization fluctuations $L_C^P \sim \sqrt{g}$ can be introduced in a conventional way (see e.g. Ref.[18]). Note, that correlation lengths $L_C^{\Phi}$ and $L_C^{P}$ determine the width of the AFD and FE domain walls [18]. Since the correlation function depends both on the wave vector of fluctuations and their orientation in **r**-space, it is anisotropic. In other words, the correlation function of tilt (or polarization) fluctuation is the second rank tensor. **Table II** lists the analytical expressions and numerical values of correlation lengths for the tilt and polarization vector components. It is seen that $L_C^{\Phi}$ varies significantly at the 180° domain wall for the different tilts (from 3.72 Å for $\Phi_1$ to 0.78 Å $\Phi_3$), and $L_C^{P}$ varies from 6.20 Å for $P_1$ to 2.38 Å to $P_2$ components. The tilt and polarization changes at 109° domain wall behave as in the hypothetic "isotropic" ferroic, namely $L_C^{\Phi} = 1.75$ Å for both tilt components and $L_C^{P} = 1.75$ Å for both polarizations. Contrary, only the components $\Phi_3$ and $P_3$ vary across the 71° domain wall, at that $L_C^{\Phi} = 1.75$ Å for $\Phi_3$ and $L_C^{P} = 2.38$ Å for $P_3$.

**Table II.** AFD and FE order parameter correlation lengths $L_C^{\Phi}$ and $L_C^{P}$ in BFO

| **Order parameter** | **Type of the uncharged domain wall** | | |
|---|---|---|---|
| | **180°** | **109°** | **71°** |
| $\Phi_1$ | $L_C^{\Phi} = \sqrt{\frac{v_{11} + v_{12} + 2v_{44}}{-4b_1}}$ =3.72 Å | Non applicable, since $\Phi_1 \approx const$ | Non applicable, since $\Phi_1 \approx const$ |
| $\Phi_2$ | $L_C^{\Phi} = \sqrt{\frac{v_{44}}{-2b_1}}$ =1.75 Å | $L_C^{\Phi} = \sqrt{\frac{v_{44}}{-2b_1}}$ =1.75 Å | Non applicable, since $\Phi_2 \approx const$ |
| $\Phi_3$ | $L_C^{\Phi} = \sqrt{\frac{v_{11} - v_{12}}{-4b_1}}$ =0.78 Å | $L_C^{\Phi} = \sqrt{\frac{v_{44}}{-2b_1}}$ =1.75 Å | $L_C^{\Phi} = \sqrt{\frac{v_{44}}{-2b_1}}$ =1.75 Å |
| $P_1$ | $L_C^{P} = \sqrt{\frac{g_{11} + g_{12} + 2g_{44}}{-4a_1}}$ =6.20 Å | Non applicable, since $P_1 \approx const$ | Non applicable, since $P_1 \approx const$ |
| $P_2$ | $L_C^{P} = \sqrt{\frac{g_{44}}{-2a_1}}$ =2.38 Å | $L_C^{P} = \sqrt{\frac{g_{44}}{-2a_1}}$ =2.38 Å | Non applicable, since $P_2 \approx const$ |
| $P_3$ | $L_C^{P} = \sqrt{\frac{g_{11} - g_{12}}{-4a_1}}$ =4.87 Å | $L_C^{P} = \sqrt{\frac{g_{44}}{-2a_1}}$ =2.38 Å | $L_C^{P} = \sqrt{\frac{g_{44}}{-2a_1}}$ =2.38 Å |

### B. Meandering 180-degree AFD-FE domain walls

Using FEM of the AFD and FE properties of strain-free thin BFO films, we further observe that the conventional 180° domains of bulk rhombohedral AFD-FE phase [see **Fig.1(a)**] are separated by the zig-zag like meandering domain walls, which in fact contain thin AFD-FE domain regions of lower monoclinic symmetry and different parity [see **Fig. 5**].

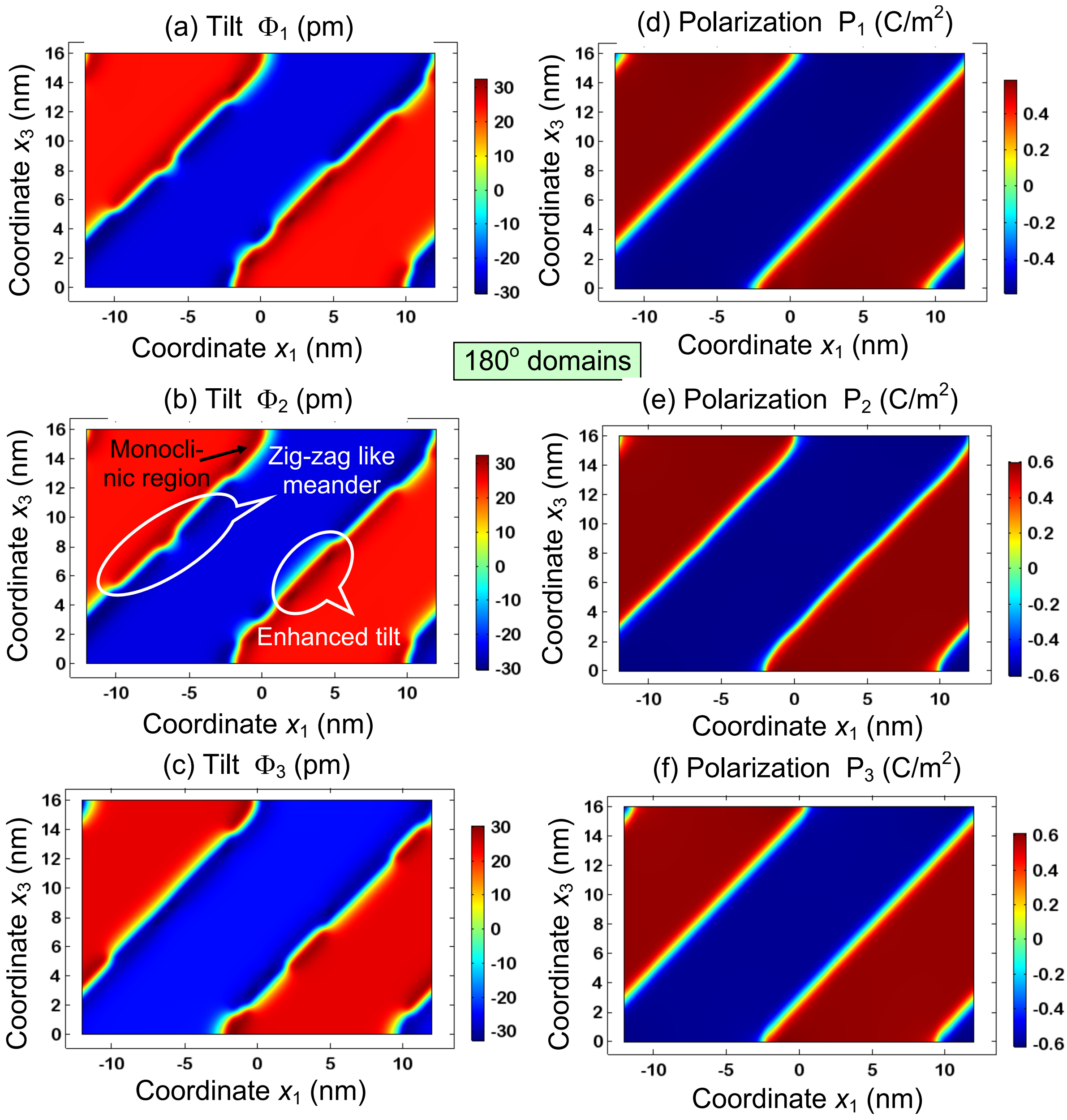

**FIGURE 5**. Distribution of the tilt $\Phi_i$ **(a)-(c)** and polarization $P_i$ **(d)-(f)** components calculated for the case of 180°-domains in a 16-nm BFO film. . Gradient coefficient $v_{44}$=2.5×10$^{10}$ J/ m$^3$, T=300 K, other parameters are listed in **Table I.**

The contrast of the monoclinic domains is determined by the magnitude of the tilt components $\Phi_i$ ; and it is higher in the vicinity of the meandering walls in comparison with the contrast in the centre of the 180° domain [see dark-red and dark-blue regions near meandering boundaries in **Fig. 5(a)-(c)**]. Surprisingly, neither curvature nor enhanced contrast is inherent to the FE component of the 180° domain boundaries [see straight incline domain boundaries with gradually changing color from red to blue in **Fig. 5(d)-(f)**]. Actually, the contrast enhancement in the meandering regions [marked by the ellipse in **Fig. 5(b)**] does not correspond to the bulk rhombohedral phase and represents itself the domains of lower monoclinic symmetry with $\Phi_1 \neq \Phi_2 \neq \Phi_3$ imposed on the 180° AFD-FE domains in the rhombohedral phase.

The influence of the tilt gradient coefficient $v_{44}$ on the domain structure could be seen from **Figs. 6-7.** Meandering AFD domain walls broaden significantly and decrease their curvature with an increase of $v_{44}$ by a factor of 10. In addition to significant broadening, a visible asymmetry of the wall profile appears with an increase of $v_{44}$ by a factor of 100. As one could see from the figures, the maximal deviation of the tilt from bulk value is dependent on $v_{44}$, but polarization profiles are almost independent on this parameter. Thus, we conclude that the appearance of meandering walls and low symmetry phases is conditioned by the decrease of the tilts gradient energy. If $v_{44}$ is sufficiently small, the energy increase associated with the AFD wall bending is less than the energy decrease associated with the terms proportional to $b_{ij}\Phi_i^2\Phi_j^2$.

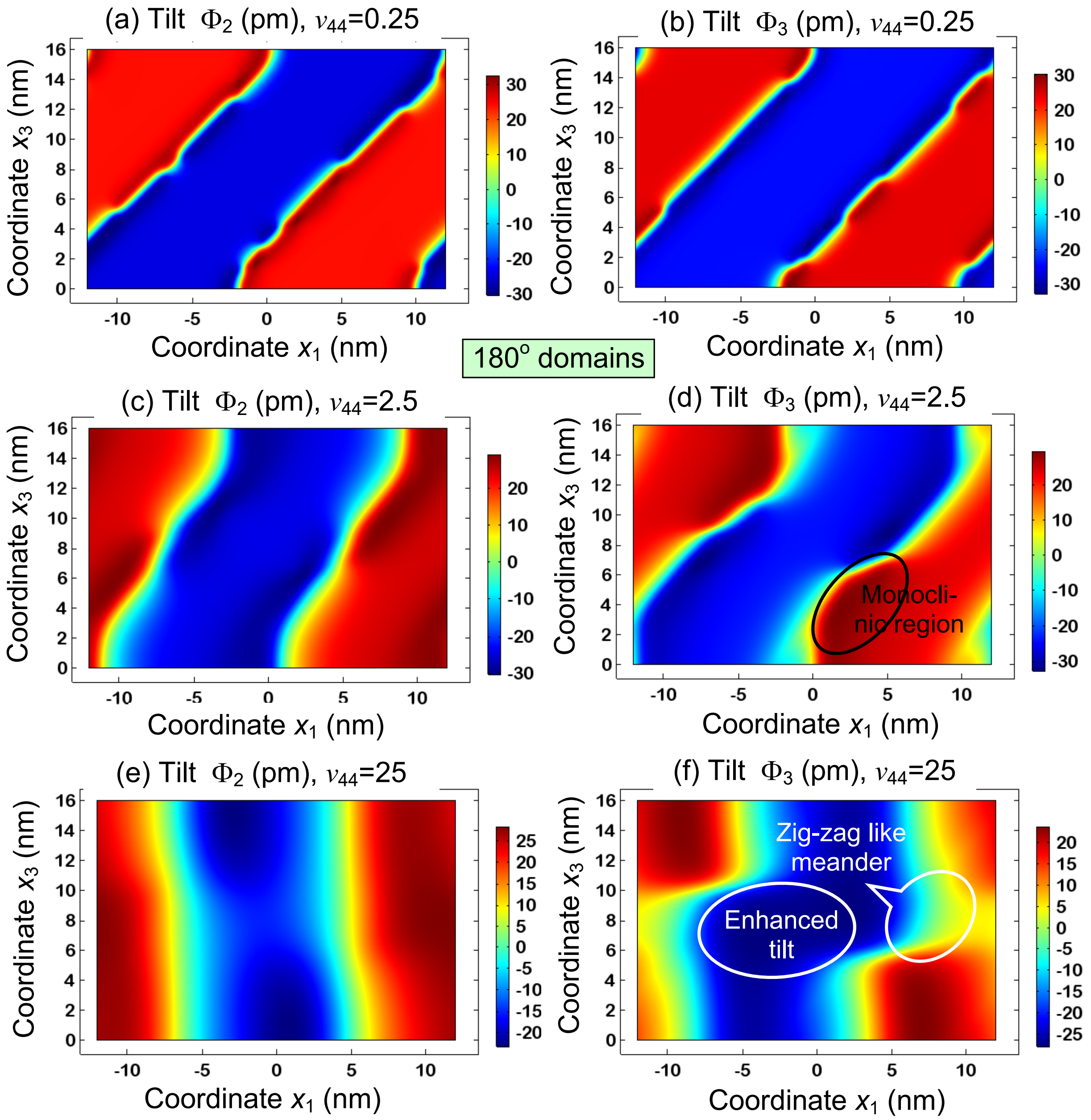

**FIGURE 6**. Distribution of the tilt components $\Phi_2$ **(a, c, e)** and $\Phi_3$ **(b, d, f)** calculated for the case of 180°-domains in a thin BFO film. The gradient coefficient $v_{44}$=2.5×10$^{10}$ J/m$^3$ **(a, b)**, 2.5×10$^{11}$ J/m$^3$ **(c, d)**, and 2.5×10$^{12}$ J/m$^3$ **(e, f).** T=300 K, other parameters are listed in **Table I.**

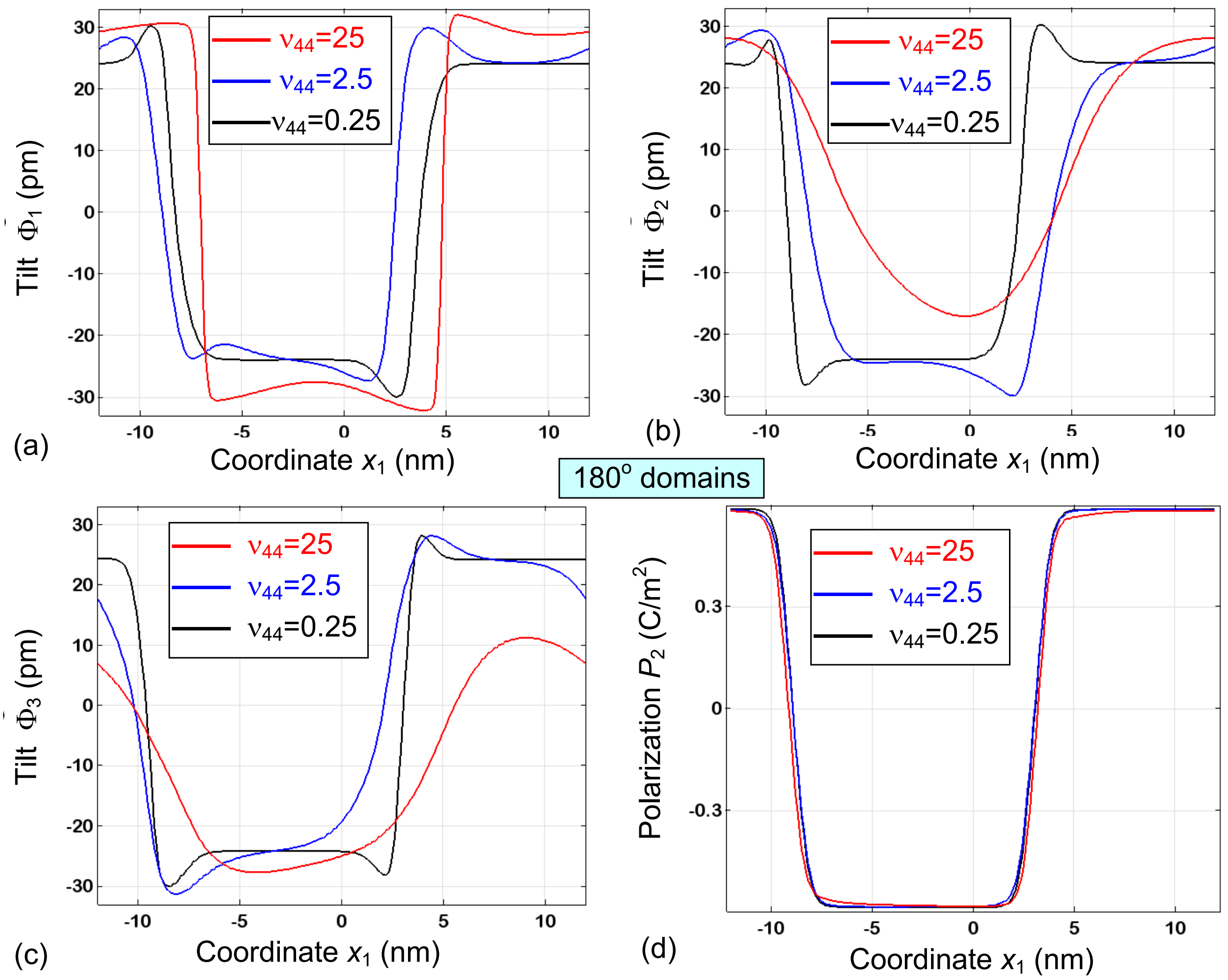

**FIGURE 7**. Profiles of the tilt components $\Phi_i$ **(a, b, c)**, and polarization $P_2$ **(d)** calculated for the case of 180°-domains in the middle of 16-nm BFO film ($x_3 = h/2$) at room temperature. The gradient coefficient $\nu_{44}$=0.25×10$^{11}$ J/m$^3$ (black curves), 2.5×10$^{11}$ J/m$^3$ (blue curves) and 25×10$^{11}$ J/m$^3$ (red curves); T=300 K, other parameters are listed in **Table I.**

### C. Meandering 109-degree AFD-FE domain walls

Rhombohedral 109° domains correspond to the case when two components of vectorial order parameter changes its sign when crossing the wall plane [see **Fig. 1(b)**]. These are the components $\Phi_2$, $\Phi_3$ and $P_2$, $P_3$, respectively for the 109° domains in BFO. These domains are separated by the AFD meandering domain walls, which in fact contain thin domains of lower symmetry [see **Figs. 8(b,c)**]. Enhanced contrast is also inherent to the FE component at the domain boundaries [see the boundaries with gradually changing color from red to blue in **Figs. 8(a)-(c)**]. Actually, the contrast enhancement in the meandering regions does not correspond to the bulk rhombohedral R3c phase.

There are the domains of lower symmetry with $\Phi_1 \neq \Phi_2 \neq \Phi_3$ and $P_1 \neq P_2 \neq P_3$ imposed on the twin boundaries in the rhombohedral phase.

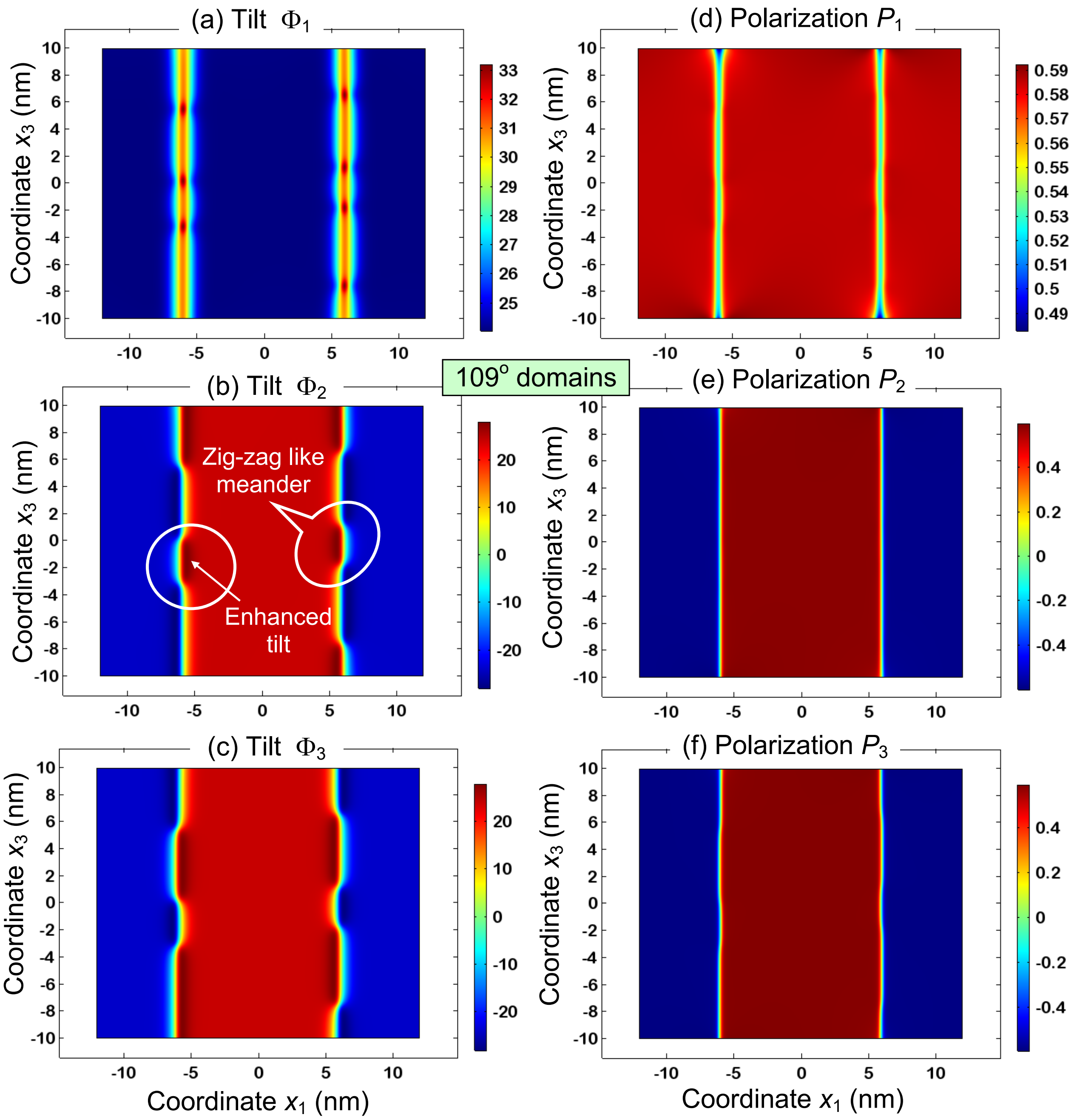

**FIGURE 8**. Distribution of the AFD order parameter $\Phi_i$ **(a)-(c)** and polarization $P_i$ **(d)-(f)** calculated for the case of 109°-domains in a thin BFO film. T=300 K, BFO parameters are listed in **Table I.**

Similarly to the case of the low symmetry phases appearing the vicinity of the meandering 180° AFD-FE domain walls (considered in section III.B) we made sure that the appearance of low symmetry domains at 109° domain walls does not steam from the spatial confinement or imperfect screening of spontaneous polarization, electrostrictive or flexoelectric couplings, but rather from the interplay between the gradient of the oxygen tilt and polarization components at the domain walls. Indeed, the influence on the tilt gradient coefficient value is shown in **Fig. 9.** Distributions of order parameters in the central part of the film, corresponding to **Fig. 9,** are shown in **Fig. 10.**

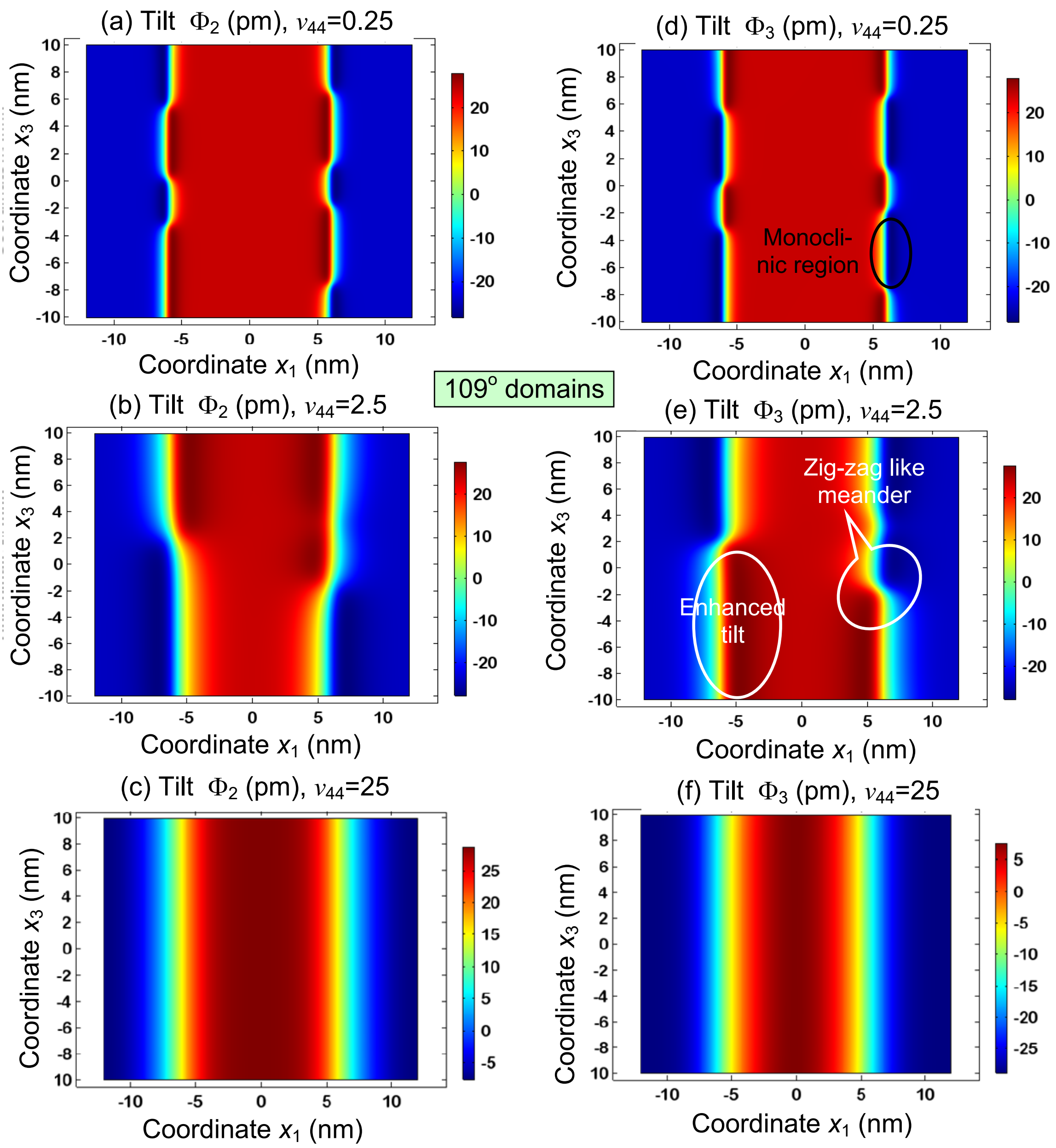

**FIGURE 9**. Distributions of the tilt components $\Phi_2$ **(a)-(c)** and $\Phi_3$ **(d)-(f)** calculated for the case of 109°-domains in a thin BFO film. The gradient coefficient $v_{44}$=0.25×10$^{11}$ J/m$^3$ **(a, e)**, 0.5×10$^{11}$ J/m$^3$ **(b, f)**, 1×10$^{11}$ J/m$^3$ **(c, g)** and 2×10$^{11}$ J/m$^3$ **(d, h).** T=300 K, BFO parameters are listed in **Table I.**

One could see two tendencies with increase of $\nu_{44}$. The first tendency is an obvious increase of domain wall width (proportionally to $\sqrt{\nu_{44}}$) and the second one is the decrease of the meandering walls density and curvature, which separate monoclinic regions. As one could see from **Fig. 10**, the amplitude of the tilt deviation from the bulk value is independent on the gradient coefficient $v_{44}$. However the low symmetry region occupies the central part of BFO film for the high values of $v_{44}$. The regions are characterized by the different amplitudes of the tilt components near and far from the 109° domain walls.

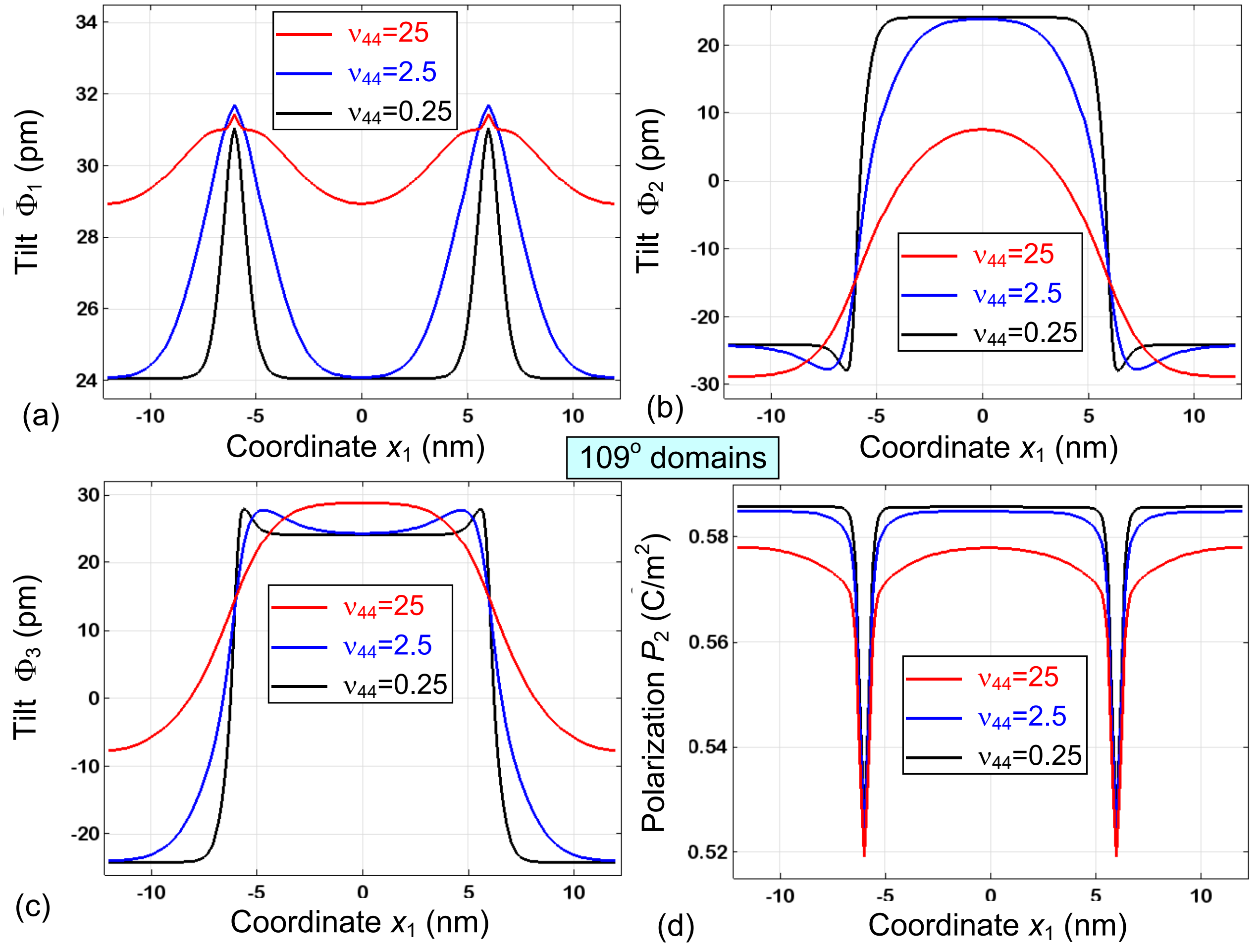

**FIGURE 10**. Profiles of the tilt $\Phi_2$ **(a, b, c)**, and polarization $P_2$ **(d)** calculated for the case of 109°-domains in a 16-nm BFO film at room temperature. The gradient coefficient $v_{44}$=0.25×10$^{11}$ J/m$^3$ (black), 2.5×10$^{11}$ J/m$^3$ (blue) and 25×10$^{11}$ J/m$^3$ (red); $x_3 = h/2$; T=300 K, BFO parameters are listed in **Table I.**

### D. 71-degree AFD-FE domain walls

Bulk 71° domains correspond to the case when only one component of the tilt and polarization changes its sign when crossing the wall plane [see **Fig.1(c)**]. It is $\Phi_2$ and $P_2$ in the considered R3c phase of BFO. The results of calculations are presented in **Figs. 11-12.** AFD-FE 71° domain walls are almost straight except for the slight bending at the surface (see **Fig. 11**). The profiles of tilt and polarization components broaden with increase of the tilt gradient coefficient $\nu_{44}$ (compare black, blue and red curves in **Figs. 12**). The tilt amplitude remains the same and polarization amplitude decreases with $\nu_{44}$ increase.

Slight bending of the FE domain wall at the surface is determined by internal electric field that has nonzero out-of-plane component near the surface (see **Fig. S.3b**). The in-plane component of the field is maximal at the domain walls far from the surface (see **Fig. S.3a**). The effect is caused by the variation of in-plane component of polarization $P_1$ perpendicular to the wall due to the coupling with other components of polarization and tilt (see **Figs. 11d**). However, at the surface the electric field should be perpendicular to it, hence the component $E_1$ tends to zero here (see **Fig. S.3a**) and the domain wall - surface junction acts as a source of a stray electric field, causing the wall bending and broadening in this region.

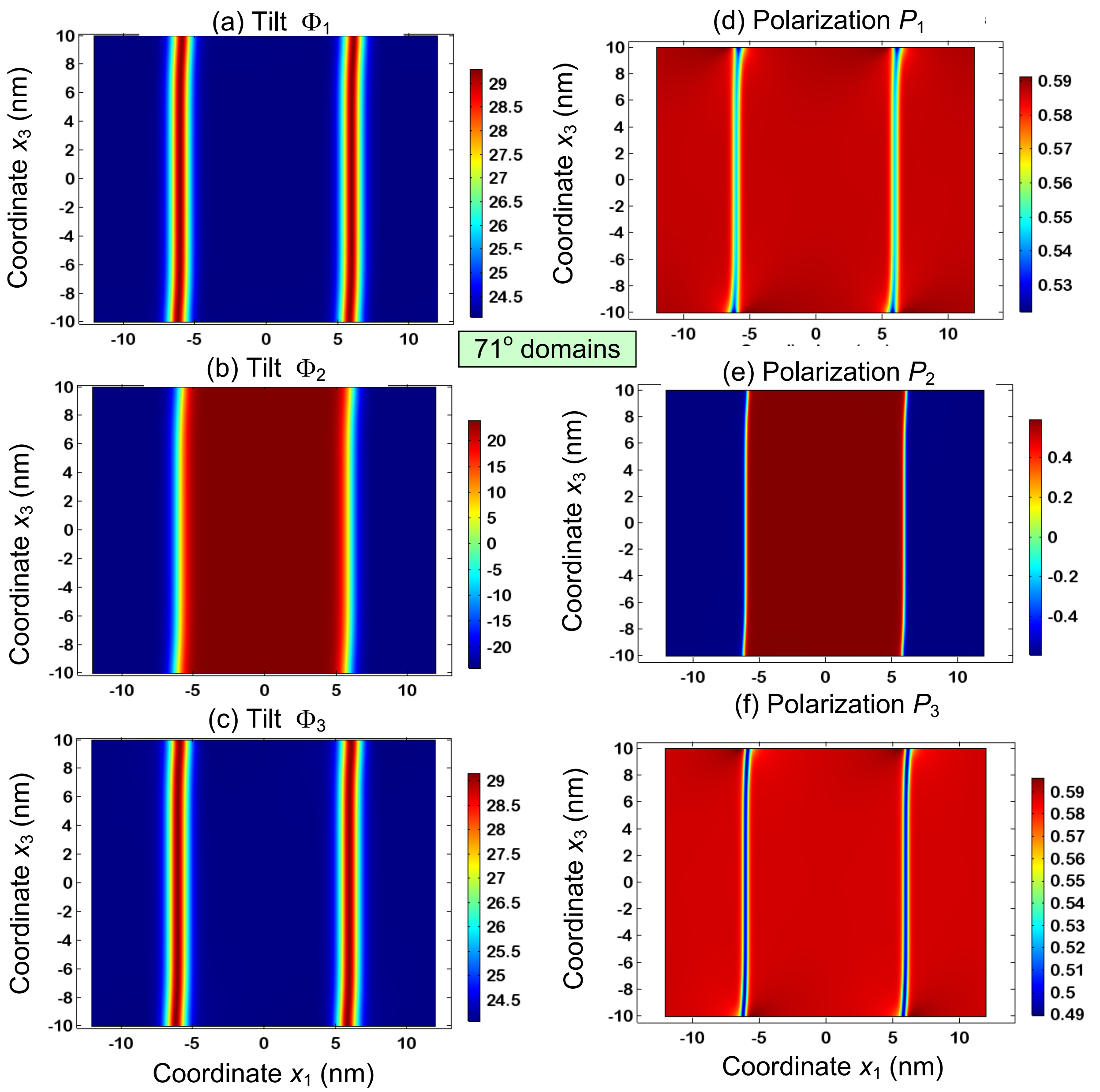

**FIGURE 11**. Distributions of the tilt $\Phi_i$ **(a)-(c)** and polarization $P_i$ **(d)-(f)** components calculated for the case of 71°-domains in a 16-nm BFO film. Gradient coefficient $v_{44}$=0.25×10$^{11}$ J/m$^3$, T=300 K, BFO parameters are listed in **Table I.** Initial distribution corresponds to 71° domains with [100] walls.

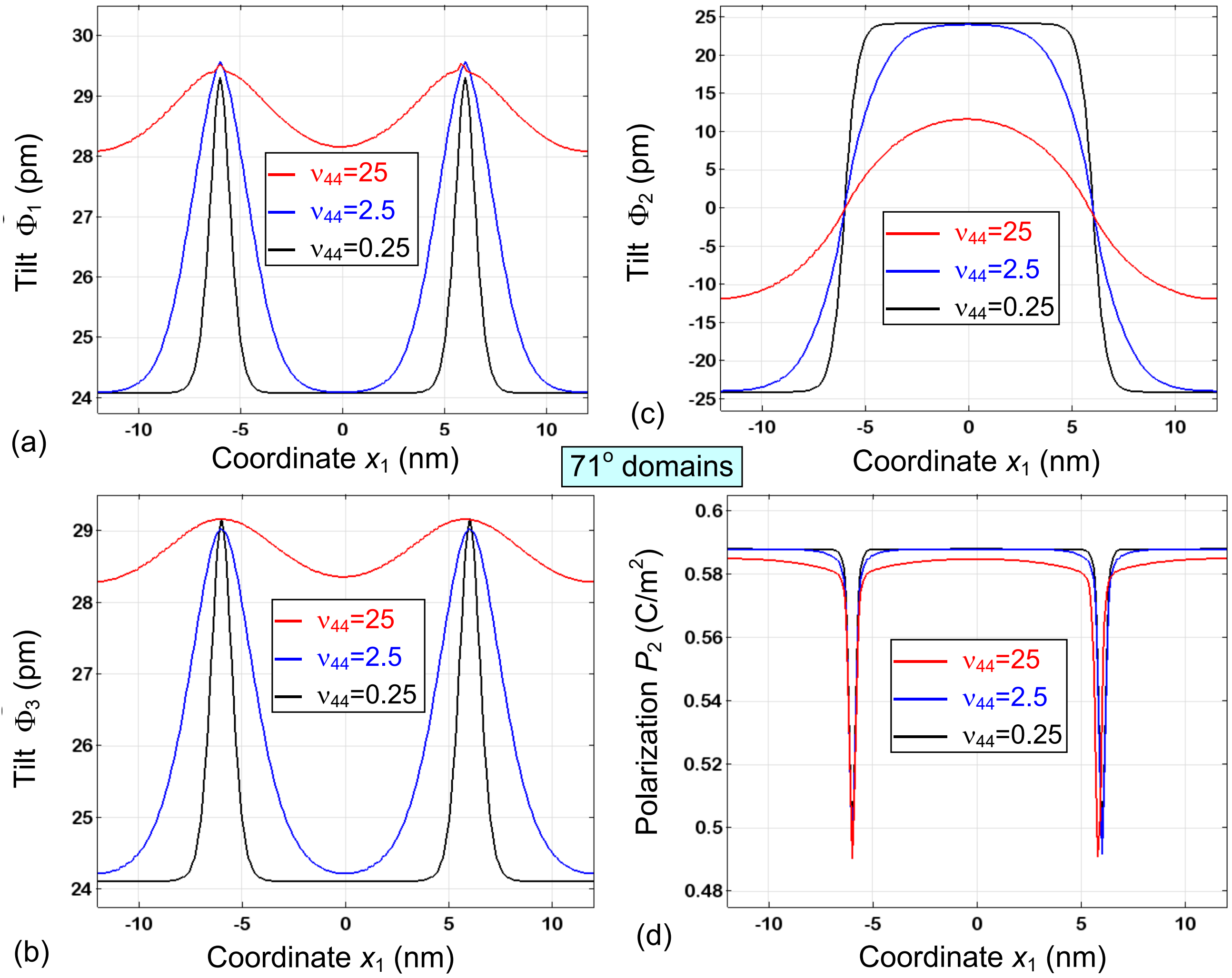

**FIGURE 12**. Profiles of tilts $\Phi_1$ **(a),** $\Phi_3$ **(b),** $\Phi_2$ **(c)**, and polarization component $P_3$ **(d)** at the surface of thin BFO film. The gradient coefficient $v_{44}$=0.25×10$^{11}$ J/m$^3$ (black curves), 2.5×10$^{11}$ J/m$^3$ (blue curves) and 25×10$^{11}$ J/m$^3$ (red curves). T=300 K, BFO parameters are listed in **Table I.**

## IV. CONCLUSIONS

Using LGD-approach we revealzig-zag like meandering AFD-FE domain walls in BFO. These walls typically separate the regions with unusually low monoclinic symmetry. It appeared that the origin of the meandering AFD-FE is conditioned by the decrease of the tilts gradient energy.

Moreover, the origin of the meandering walls does not steam from incomplete polarization screening in thin BFO films, electrostrictive or flexoelectric coupling. The spatial confinement delineates the appropriate boundary conditions for the oxygen tilt and polarization components at the film surfaces, but its existence is not critical for the meandering walls appearance and their zig-zag like instability. The values of the gradient energy coefficients for the oxygen tilt appeared critical to initiate the morphological changes of the 180° and 109° uncharged domain walls towards zig-zag meandering. Zig-zag instability appears for small gradient energies, while the walls become straight and broaden at higher gradients. Uncharged 71° walls are always straight and their width increases with increasing of the tilt gradient coefficient.

Hence we predicted previously unexplored type of the gradient-driven morphological phase transition taking place at the AFD-FE domain walls in multiferroics.

**Acknowledgements**

Authors are very grateful to Prof. Xiuliang Ma for stimulating discussions and very useful remarks. A.N.M. work has received funding from the European Union's Horizon 2020 research and innovation programme under the Marie Skłodowska-Curie grant agreement No 778070, and partially supported by the National Academy of Sciences of Ukraine (project No. 0117U002612, No. 0118U003375) and by the Program of Fundamental Research of the Department of Physics and Astronomy of the National Academy of Sciences of Ukraine (project No. 0117U000240). S.V.K. and C.T.N. was supported by the Center for Nanophase Materials Sciences, sponsored by the Division of User Facilities, Basic Energy Sciences, US Department of Energy.

**Authors' contribution.** E.A.E. wrote the codes, performed numerical calculations and prepared figures. A.N.M. generated research idea, stated the problem, derived analytical results and wrote the manuscript draft. E.A.E. and A.N.M. contributed equality to the results interpretation. C.T.N. and S.V.K. worked on the results discussion and manuscript improvement.

## SUPPLEMENTARY MATERIALS

## APPENDIX A

### Evident form of the free energy in BFO

Explicit form of different contributions to the free energy are presented below for cubic m3m symmetry parent phase. AFD contribution is:

$$
\begin{aligned}
\Delta G_{AFD} &= b_1\left(\Phi_1^2+\Phi_2^2+\Phi_3^2\right)+b_{11}\left(\Phi_1^4+\Phi_2^4+\Phi_3^4\right)+b_{12}\left(\Phi_1^2\Phi_2^2+\Phi_1^2\Phi_3^2+\Phi_3^2\Phi_2^2\right)+\\
& b_{111}\left(\Phi_1^6+\Phi_2^6+\Phi_3^6\right)+b_{112}\left(\Phi_1^2\left(\Phi_2^4+\Phi_3^4\right)+\Phi_1^4\left(\Phi_2^2+\Phi_3^2\right)+\Phi_2^2\Phi_3^4+\Phi_2^2\Phi_3^2\right)+b_{123}\Phi_1^2\Phi_2^2\Phi_3^2+\\
&+v_{11}\left(\left(\frac{\partial\Phi_1}{\partial x_1}\right)^2+\left(\frac{\partial\Phi_2}{\partial x_2}\right)^2+\left(\frac{\partial\Phi_3}{\partial x_3}\right)^2\right)+v_{12}\left(\frac{\partial\Phi_1}{\partial x_1}\frac{\partial\Phi_2}{\partial x_2}+\frac{\partial\Phi_2}{\partial x_2}\frac{\partial\Phi_3}{\partial x_3}+\frac{\partial\Phi_1}{\partial x_1}\frac{\partial\Phi_3}{\partial x_3}\right)+\\
& v_{44}\left(\left(\frac{\partial\Phi_1}{\partial x_2}\right)^2+\left(\frac{\partial\Phi_1}{\partial x_3}\right)^2+\left(\frac{\partial\Phi_2}{\partial x_1}\right)^2+\left(\frac{\partial\Phi_2}{\partial x_3}\right)^2+\left(\frac{\partial\Phi_3}{\partial x_1}\right)^2+\left(\frac{\partial\Phi_3}{\partial x_2}\right)^2\right)+\\
&+v'_{44}\left(\frac{\partial\Phi_1}{\partial x_2}\frac{\partial\Phi_2}{\partial x_1}+\frac{\partial\Phi_1}{\partial x_3}\frac{\partial\Phi_3}{\partial x_1}+\frac{\partial\Phi_2}{\partial x_3}\frac{\partial\Phi_3}{\partial x_2}\right)
\end{aligned}
\qquad (A.1)
$$

Order parameter $\Phi_i$ is the vector of tilt of oxygen octahedral groups with components (*i*=1, 2, 3).

FE contribution is:

$$
\begin{aligned}
\Delta G_{FE} &= a_1\left(P_1^2+P_2^2+P_3^2\right)+a_{11}\left(P_1^4+P_2^4+P_3^4\right)+a_{12}\left(P_1^2P_2^2+P_1^2P_3^2+P_2^2P_3^2\right)\\
&+a_{111}\left(P_1^6+P_2^6+P_3^6\right)+a_{112}\left(P_1^2\left(P_2^4+P_3^4\right)+P_1^4\left(P_2^2+P_3^2\right)+P_2^2P_3^4+P_2^4P_3^2\right)+a_{123}P_1^2P_2^2P_3^2\\
&+g_{11}\left(\left(\frac{\partial P_1}{\partial x_1}\right)^2+\left(\frac{\partial P_2}{\partial x_2}\right)^2+\left(\frac{\partial P_3}{\partial x_3}\right)^2\right)+g_{12}\left(\frac{\partial P_1}{\partial x_1}\frac{\partial P_2}{\partial x_2}+\frac{\partial P_2}{\partial x_2}\frac{\partial P_3}{\partial x_3}+\frac{\partial P_1}{\partial x_1}\frac{\partial P_3}{\partial x_3}\right)+\\
&+g_{44}\left(\left(\frac{\partial P_1}{\partial x_2}\right)^2+\left(\frac{\partial P_1}{\partial x_3}\right)^2+\left(\frac{\partial P_2}{\partial x_1}\right)^2+\left(\frac{\partial P_2}{\partial x_3}\right)^2+\left(\frac{\partial P_3}{\partial x_1}\right)^2+\left(\frac{\partial P_3}{\partial x_2}\right)^2\right)+\\
&+g'_{44}\left(\frac{\partial P_1}{\partial x_2}\frac{\partial P_2}{\partial x_1}+\frac{\partial P_1}{\partial x_3}\frac{\partial P_3}{\partial x_1}+\frac{\partial P_2}{\partial x_3}\frac{\partial P_3}{\partial x_2}\right)-P_iE_i-\frac{\varepsilon_0\varepsilon_b}{2}E_i^2
\end{aligned}
\qquad (A.2)
$$

Polarization vector $P_i$ has all three components for *i*=1, 2, 3.

Contribution for the biquadratic coupling between polarization and tilt

$$
\begin{aligned}
\Delta G_{BQC} &= \xi_{11}\left(\Phi_1^2P_1^2+\Phi_2^2P_2^2+\Phi_3^2P_3^2\right)+\xi_{12}\left(\left(\Phi_2^2+\Phi_3^2\right)P_1^2+\left(\Phi_1^2+\Phi_3^2\right)P_2^2+\left(\Phi_1^2+\Phi_2^2\right)P_3^2\right)\\
&+\xi_{44}\left(\Phi_1\Phi_2P_1P_2+\Phi_1\Phi_3P_1P_3+\Phi_2\Phi_3P_2P_3\right)
\end{aligned}
. \qquad (A.3)
$$

Electrostriction and rotostriction contributions to the free energy are

$$
\begin{aligned}
\Delta G_{STR} = & -Q_{11}\left(\sigma_{11}P_1^2+\sigma_{22}P_2^2+\sigma_{33}P_3^2\right)\\
& -Q_{12}\left((\sigma_{22}+\sigma_{33})P_1^2+(\sigma_{11}+\sigma_{33})P_2^2+(\sigma_{22}+\sigma_{11})P_3^2\right)\\
& -Q_{44}\left(\sigma_{12}P_1P_2+\sigma_{13}P_1P_3+\sigma_{23}P_2P_3\right)\\
& -R_{11}\left(\sigma_{11}\Phi_1^2+\sigma_{22}\Phi_2^2+\sigma_{33}\Phi_3^2\right)-R_{12}\left((\sigma_{22}+\sigma_{33})\Phi_1^2+(\sigma_{11}+\sigma_{33})\Phi_2^2+(\sigma_{22}+\sigma_{11})\Phi_3^2\right)\\
& -R_{44}\left(\sigma_{12}\Phi_1\Phi_2+\Phi_1\Phi_3+_{23}\Phi_2\Phi_3\right)
\end{aligned}
\tag{A.4}
$$

Flexoelectric effect contribution is

$$
\begin{aligned}
\Delta G_{\mathrm{FL}} = & -F_{11}\left(\sigma_{11}\frac{\partial P_1}{\partial x_1}+\sigma_{22}\frac{\partial P_2}{\partial x_2}+\sigma_{33}\frac{\partial P_3}{\partial x_3}\right)\\
& -F_{12}\left((\sigma_{22}+\sigma_{33})\frac{\partial P_1}{\partial x_1}+(\sigma_{11}+\sigma_{33})\frac{\partial P_2}{\partial x_2}+(\sigma_{11}+\sigma_{22})\frac{\partial P_3}{\partial x_3}\right)\\
& -F_{44}\left(\sigma_{12}\left(\frac{\partial P_1}{\partial x_2}+\frac{\partial P_2}{\partial x_1}\right)+\sigma_{13}\left(\frac{\partial P_1}{\partial x_3}+\frac{\partial P_3}{\partial x_1}\right)+\sigma_{23}\left(\frac{\partial P_2}{\partial x_3}+\frac{\partial P_3}{\partial x_2}\right)\right)
\end{aligned}
\tag{A.5}
$$

Elastic "self-energy" is

$$
\Delta G_{\sigma} = -\frac{s_{11}}{2}\left(\sigma_{11}^2+\sigma_{22}^2+\sigma_{33}^2\right)-s_{12}\left(\sigma_{11}\sigma_{22}+\sigma_{22}\sigma_{33}+\sigma_{11}\sigma_{33}\right)-\frac{s_{44}}{2}\left(\sigma_{12}^2+\sigma_{23}^2+\sigma_{13}^2\right)
\tag{A.6}
$$

Total free energy has the following form

$$
G=\int_V\left(\Delta G_{AFD}+\Delta G_{FE}+\Delta G_{BQC}+\Delta G_{STR}+\Delta G_{FL}+\Delta G_{\sigma}\right)dv+\int_S\left(\Delta G_{AFD}+\Delta G_{FE}\right)dS
\tag{A.7}
$$

The Euler-Lagrange equations for polarization and tilt components, $P_i$ and $\Phi_i$, can be obtained by the minimization of the free energy (A.7) as follows

$$
\begin{aligned}
-\Gamma_P\frac{\partial P_1}{\partial t} = & \,2P_1\left(a_1-Q_{12}(\sigma_{22}+\sigma_{33})-Q_{11}\sigma_{11}\right)-Q_{44}\left(\sigma_{12}P_2+\sigma_{13}P_3\right)\\
& +\left(\zeta_{11}\Phi_1^2+\zeta_{12}\left(\Phi_2^2+\Phi_3^2\right)\right)2P_1+\zeta_{44}\Phi_1\left(\Phi_2P_2+\Phi_3P_3\right)\\
& +4a_{11}P_1^3+2a_{12}P_1\left(P_2^2+P_3^2\right)+6a_{111}P_1^5+2a_{112}P_1\left(P_2^4+2P_1^2P_2^2+P_3^4+2P_1^2P_3^2\right)+2a_{112}P_1P_2^2P_3^2\\
& -g_{11}\frac{\partial^2P_1}{\partial x_1^2}-g_{44}\left(\frac{\partial^2P_1}{\partial x_2^2}+\frac{\partial^2P_1}{\partial x_3^2}\right)-\left(g'_{44}+g_{12}\right)\frac{\partial^2P_2}{\partial x_2\partial x_1}-\left(g'_{44}+g_{12}\right)\frac{\partial^2P_3}{\partial x_3\partial x_1}\\
& +F_{11}\frac{\partial\sigma_{11}}{\partial x_1}+F_{12}\left(\frac{\partial\sigma_{22}}{\partial x_1}+\frac{\partial\sigma_{33}}{\partial x_1}\right)+F_{44}\left(\frac{\partial\sigma_{12}}{\partial x_2}+\frac{\partial\sigma_{13}}{\partial x_3}\right)-E_1
\end{aligned}
$$

(A.8a)

$$\begin{aligned}
-\Gamma_P \frac{\partial P_2}{\partial t} &= 2P_2\left(a_1 - Q_{12}(\sigma_{11}+\sigma_{33}) - Q_{11}\sigma_{22}\right) - Q_{44}\left(\sigma_{12}P_1 + \sigma_{23}P_3\right) \\
&+ \left(\zeta_{11}\Phi_2^2 + \zeta_{12}\left(\Phi_1^2+\Phi_3^2\right)\right)2P_2 + \zeta_{44}\Phi_2\left(\Phi_1 P_1+\Phi_3 P_3\right) \\
&+ 4a_{11}P_2^3 + 2a_{12}P_2\left(P_1^2+P_3^2\right) + 6a_{111}P_2^5 + 2a_{112}P_2\left(P_1^4 + 2P_2^2P_1^2 + P_3^4 + 2P_2^2P_3^2\right) + 2a_{112}P_2P_1^2P_3^2 \\
&- g_{11}\frac{\partial^2 P_2}{\partial x_2^2} - g_{44}\left(\frac{\partial^2 P_2}{\partial x_1^2}+\frac{\partial^2 P_2}{\partial x_3^2}\right) - \left(g'_{44}+g_{12}\right)\frac{\partial^2 P_1}{\partial x_2 \partial x_1} - \left(g'_{44}+g_{12}\right)\frac{\partial^2 P_3}{\partial x_3 \partial x_2} \\
&+ F_{11}\frac{\partial \sigma_{22}}{\partial x_2} + F_{12}\left(\frac{\partial \sigma_{11}}{\partial x_2}+\frac{\partial \sigma_{33}}{\partial x_2}\right) + F_{44}\left(\frac{\partial \sigma_{12}}{\partial x_1}+\frac{\partial \sigma_{23}}{\partial x_3}\right) - E_2
\end{aligned} \tag{A.8b}$$

$$\begin{aligned}
-\Gamma_P \frac{\partial P_3}{\partial t} &= 2P_3\left(a_1 - Q_{12}(\sigma_{11}+\sigma_{22}) - Q_{11}\sigma_{33}\right) - Q_{44}\left(\sigma_{13}P_1 + \sigma_{23}P_2\right) \\
&+ \left(\zeta_{11}\Phi_3^2 + \zeta_{12}\left(\Phi_1^2+\Phi_2^2\right)\right)2P_3 + \zeta_{44}\Phi_3\left(\Phi_1 P_1+\Phi_2 P_2\right) \\
&+ 4a_{11}P_2^3 + 2a_{12}P_3\left(P_1^2+P_2^2\right) + 6a_{111}P_3^5 + 2a_{112}P_3\left(P_1^4 + 2P_3^2P_1^2 + P_2^4 + 2P_2^2P_3^2\right) + 2a_{112}P_3P_1^2P_2^2 \\
&- g_{11}\frac{\partial^2 P_3}{\partial x_3^2} - g_{44}\left(\frac{\partial^2 P_3}{\partial x_1^2}+\frac{\partial^2 P_3}{\partial x_2^2}\right) - \left(g'_{44}+g_{12}\right)\frac{\partial^2 P_1}{\partial x_3 \partial x_1} - \left(g'_{44}+g_{12}\right)\frac{\partial^2 P_2}{\partial x_3 \partial x_2} \\
&+ F_{11}\frac{\partial \sigma_{33}}{\partial x_3} + F_{12}\left(\frac{\partial \sigma_{11}}{\partial x_3}+\frac{\partial \sigma_{33}}{\partial x_3}\right) + F_{44}\left(\frac{\partial \sigma_{13}}{\partial x_1}+\frac{\partial \sigma_{23}}{\partial x_2}\right) - E_3
\end{aligned} \tag{A.8c}$$

$$\begin{aligned}
-\Gamma_\Phi \frac{\partial \Phi_1}{\partial t} &= \left(b_1 - R_{12}(\sigma_{22}+\sigma_{33}) - R_{11}\sigma_{11} + \zeta_{11}P_1^2 + \zeta_{12}\left(P_2^2+P_3^2\right)\right)2\Phi_1 \\
&- R_{44}\left(\sigma_{12}\Phi_2 + \sigma_{13}\Phi_3\right) + \zeta_{44}P_1\left(\Phi_2 P_2 + \Phi_3 P_3\right) + 4b_{11}\Phi_1^3 + 2b_{12}\Phi_1\left(\Phi_2^2+\Phi_3^2\right) \\
&+ 6b_{111}\Phi_1^5 + 2b_{112}\Phi_1\left(\Phi_2^4 + 2\Phi_1^2\Phi_2^2 + \Phi_3^4 + 2\Phi_1^2\Phi_3^2\right) + 2b_{112}\Phi_1\Phi_2^2\Phi_3^2 \\
&- \nu_{11}\frac{\partial^2 \Phi_1}{\partial x_1^2} - \nu_{44}\left(\frac{\partial^2 \Phi_1}{\partial x_2^2}+\frac{\partial^2 \Phi_1}{\partial x_3^2}\right) - \left(\nu'_{44}+\nu_{12}\right)\frac{\partial^2 \Phi_2}{\partial x_1 \partial x_2} - \left(\nu'_{44}+\nu_{12}\right)\frac{\partial^2 \Phi_3}{\partial x_1 \partial x_3}
\end{aligned} \tag{A.9a}$$

$$\begin{aligned}
-\Gamma_\Phi \frac{\partial \Phi_2}{\partial t} &= \left(b_1 - R_{12}(\sigma_{11}+\sigma_{33}) - R_{11}\sigma_{22} + \zeta_{11}P_2^2 + \zeta_{12}\left(P_1^2+P_3^2\right)\right)2\Phi_2 \\
&- R_{44}\left(\sigma_{12}\Phi_1 + \sigma_{23}\Phi_3\right) + \zeta_{44}P_2\left(\Phi_1 P_1 + \Phi_3 P_3\right) + 4b_{11}\Phi_2^3 + 2b_{12}\Phi_2\left(\Phi_1^2+\Phi_3^2\right) \\
&+ 6b_{111}\Phi_2^5 + 2b_{112}\Phi_2\left(\Phi_1^4 + 2\Phi_1^2\Phi_2^2 + \Phi_3^4 + 2\Phi_2^2\Phi_3^2\right) + 2b_{112}\Phi_2\Phi_1^2\Phi_3^2 \\
&- \nu_{11}\frac{\partial^2 \Phi_2}{\partial x_2^2} - \nu_{44}\left(\frac{\partial^2 \Phi_2}{\partial x_1^2}+\frac{\partial^2 \Phi_2}{\partial x_3^2}\right) - \left(\nu'_{44}+\nu_{12}\right)\frac{\partial^2 \Phi_1}{\partial x_1 \partial x_2} - \left(\nu'_{44}+\nu_{12}\right)\frac{\partial^2 \Phi_3}{\partial x_2 \partial x_3}
\end{aligned} \tag{A.9b}$$

$$\begin{aligned}
-\Gamma_\Phi \frac{\partial \Phi_3}{\partial t} &= \left(b_1 - R_{12}(\sigma_{22}+\sigma_{11}) - R_{11}\sigma_{33} + \zeta_{11}P_3^2 + \zeta_{12}\left(P_2^2+P_1^2\right)\right)2\Phi_3 - R_{44}\left(\sigma_{23}\Phi_2 + \sigma_{13}\Phi_1\right) \\
&+ \zeta_{44}P_3\left(\Phi_2 P_2 + \Phi_1 P_1\right) + 4b_{11}\Phi_3^3 + 2b_{12}\Phi_3\left(\Phi_1^2+\Phi_2^2\right) + 6b_{111}\Phi_3^5 \\
&+ 2b_{112}\Phi_3\left(\Phi_2^4 + 2\Phi_3^2\Phi_2^2 + \Phi_1^4 + 2\Phi_1^2\Phi_3^2\right) + 2b_{112}\Phi_3\Phi_1^2\Phi_2^2 \\
&- \nu_{11}\frac{\partial^2 \Phi_3}{\partial x_3^2} - \nu_{44}\left(\frac{\partial^2 \Phi_3}{\partial x_2^2}+\frac{\partial^2 \Phi_3}{\partial x_1^2}\right) - \left(\nu'_{44}+\nu_{12}\right)\frac{\partial^2 \Phi_1}{\partial x_1 \partial x_3} - \left(\nu'_{44}+\nu_{12}\right)\frac{\partial^2 \Phi_2}{\partial x_2 \partial x_3}
\end{aligned} \tag{A.9c}$$

The general form of the boundary conditions is

$$b^{(S)}\Phi_i + v_{ijkl}\frac{\partial \Phi_k}{\partial x_l} n_j\bigg|_S = 0, \quad a^{(P)}P_i + \left(g_{ijkl}\frac{\partial P_k}{\partial x_l} - F_{klij}\sigma_{kl}\right)n_j\bigg|_S = 0 \quad (i=1, 2, 3) \qquad \text{(A.10)}$$

For the normal $n_l = \{0, 0, \pm 1\}$ to the surface $x_3 = 0, h$ Eqs.(A.10) have the following explicit form

$$b^{(S)}\Phi_1 \mp \left(\nu_{44}\frac{\partial \Phi_1}{\partial x_3} + \nu'_{44}\frac{\partial \Phi_3}{\partial x_1}\right)\bigg|_{x_3=0,h} = 0, \quad b^{(S)}\Phi_2 \mp \left(\nu_{44}\frac{\partial \Phi_2}{\partial x_3} + \nu'_{44}\frac{\partial \Phi_3}{\partial x_2}\right)\bigg|_{x_3=0,h} = 0,$$

$$b^{(S)}\Phi_3 \mp \left(\nu_{11}\frac{\partial \Phi_3}{\partial x_3} + \nu_{12}\frac{\partial \Phi_2}{\partial x_2} + \nu_{12}\frac{\partial \Phi_1}{\partial x_1}\right)\bigg|_{x_3=0,h} = 0 \qquad \text{(A.11a)}$$

$$a^{(S)}P_1 \mp \left(g_{44}\frac{\partial P_1}{\partial x_3} + g'_{44}\frac{\partial P_3}{\partial x_1} - F_{44}\sigma_{13}\right)\bigg|_{x_3=0,h} = 0, \quad a^{(S)}P_2 \mp \left(g_{44}\frac{\partial P_2}{\partial x_3} + g'_{44}\frac{\partial P_3}{\partial x_2} - F_{44}\sigma_{23}\right)\bigg|_{x_3=0,h} = 0,$$

$$a^{(S)}P_3 \mp \left(g_{11}\frac{\partial P_3}{\partial x_3} + g_{12}\frac{\partial P_2}{\partial x_2} + g_{12}\frac{\partial P_1}{\partial x_1} - F_{11}\sigma_{33} - F_{12}\left(\sigma_{11} + \sigma_{22}\right)\right)\bigg|_{x_3=0,h} = 0 \qquad \text{(A.11b)}$$

Note that all derivatives $\frac{\partial}{\partial x_2}$ are zero for the 2D-problem. Following Glinka and Marton semi-microscopic model, we suggested that

$$g'_{44} + g_{12} \equiv 0 \text{ and } \nu'_{44} + \nu_{12} \equiv 0 \qquad \text{(A.12)}$$

## APPENDIX B. Supplementary figures

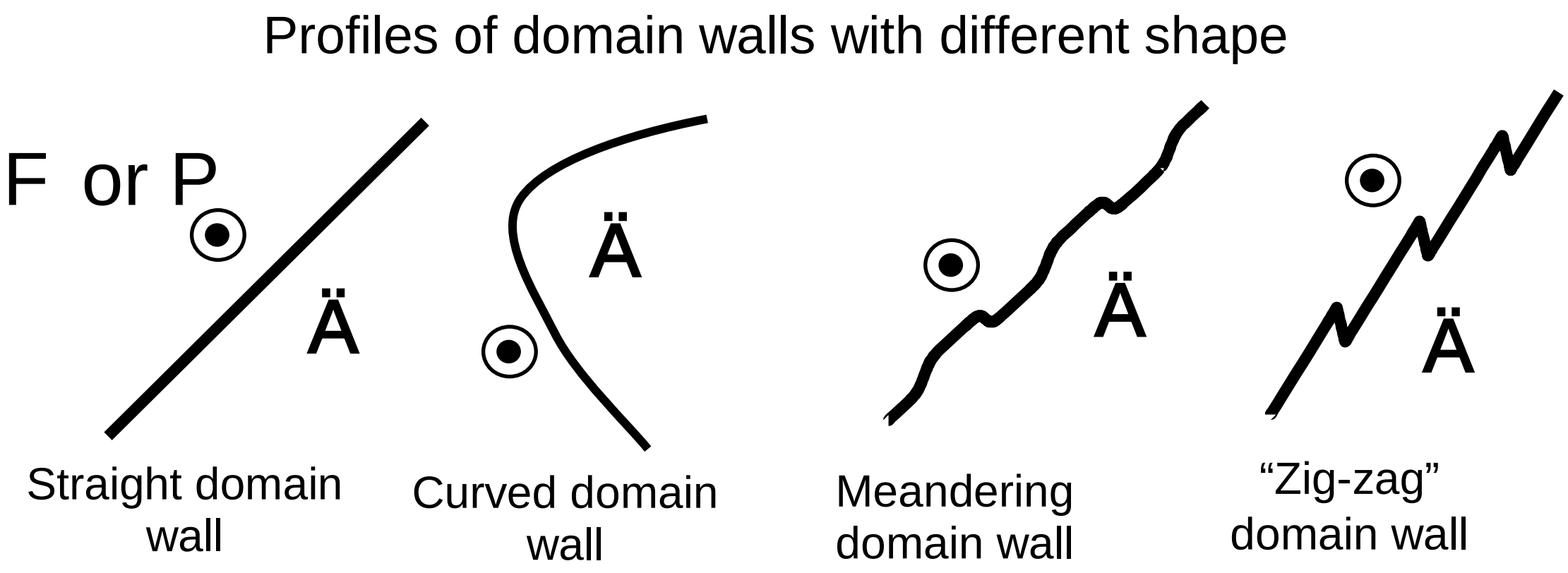

**FIGURE S1.** Schematic images of the straight, not straight or "curved", "meandering" and "zig-zag" like domain wall profiles.

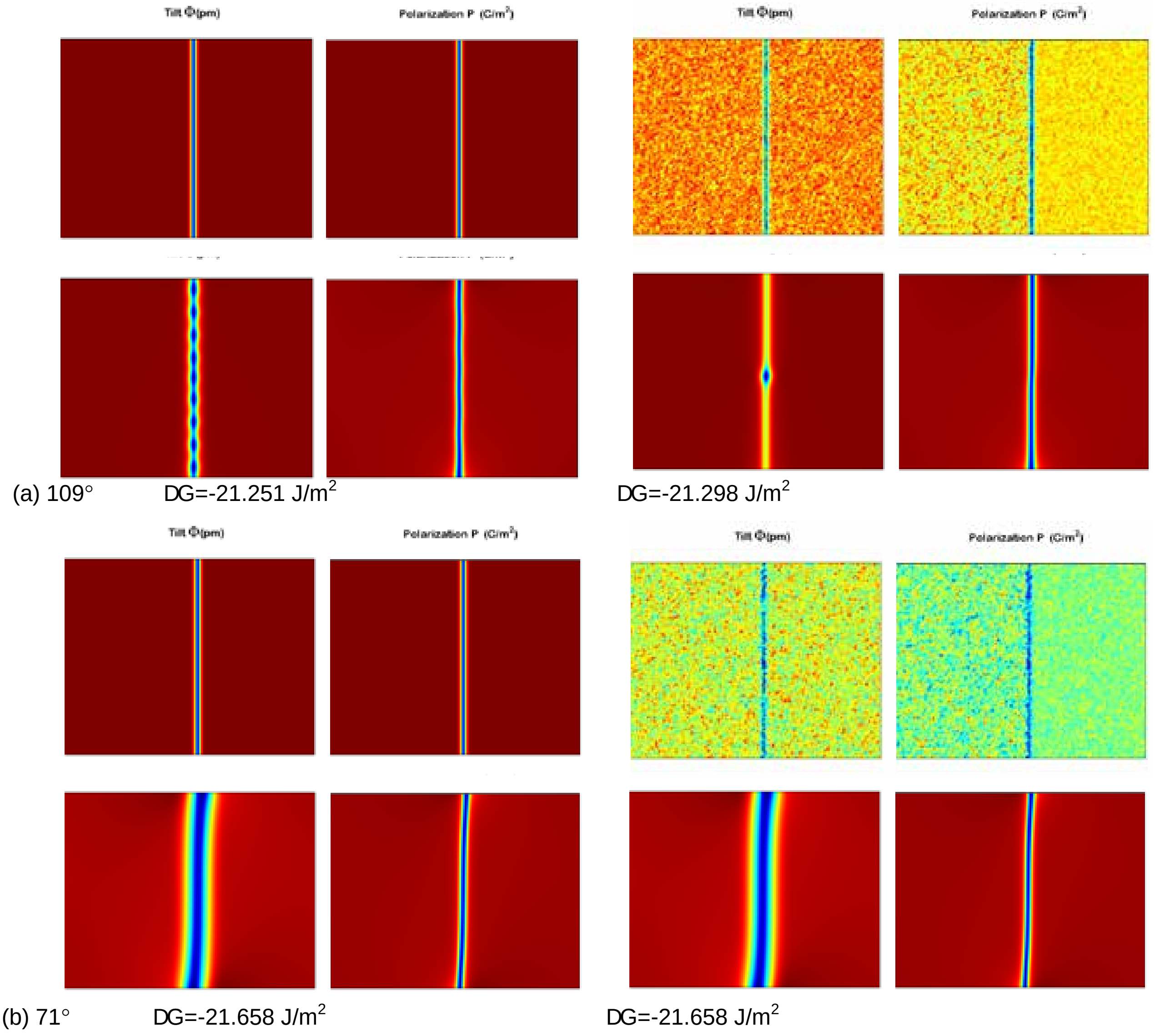

**FIGURE. S2.** Initial and final distributions of the tilt **(a, c)** and polarization **(b,d)** in a 16-nm BFO film at room temperature for the 109° and 71° domains and random seeding. Gradient coefficient $n_{44}$=0.25´ $10^{11}$ J/ $m^3$, L =0, other parameters are listed in **Table I.**

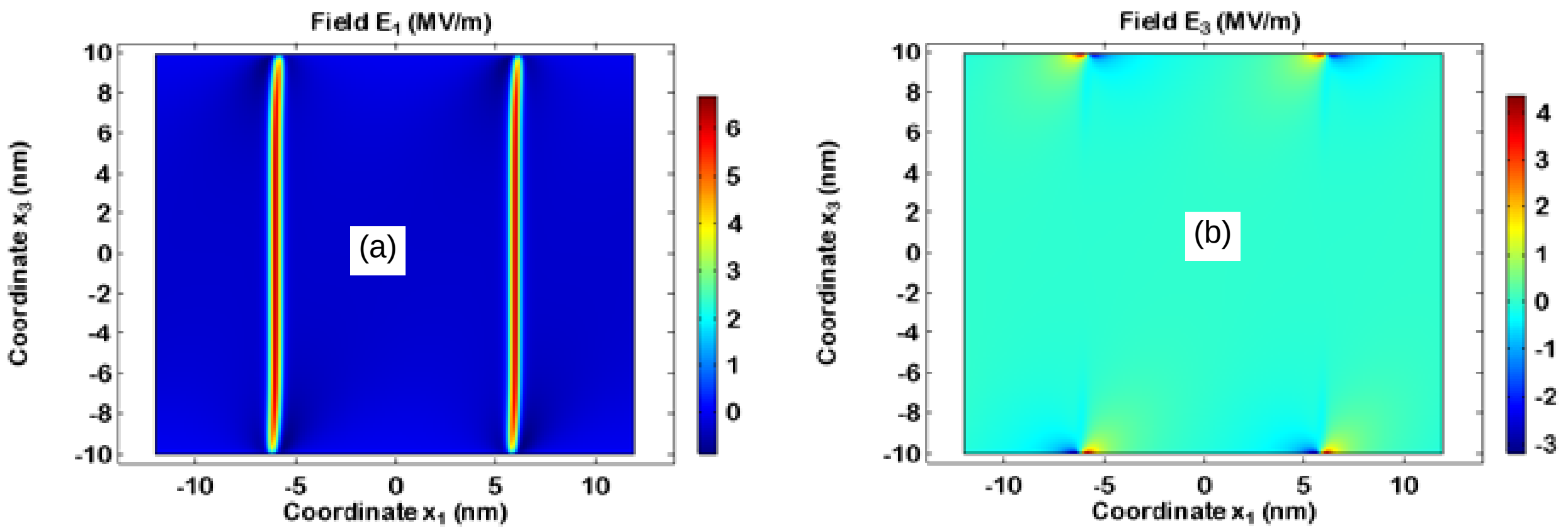

**FIGURE S3**. Distribution of the electric field components, $E_1$ **(a)** and $E_3$ **(b)** in a thin BFO film at room temperature for the case of 71° domains with [100] walls. Gradient coefficient $n_{44}$=0.25´ $10^{11}$ J/ $m^3$, other parameters are listed in **Table I.**